\documentclass[a4paper,11pt]{article}
\usepackage{jheppub}

\usepackage{tikz}
\usetikzlibrary{calc,decorations.markings,decorations.pathmorphing}
\usetikzlibrary{shapes,arrows,positioning,automata,backgrounds,calc,er,patterns}
\usepackage[compat=1.1.0]{tikz-feynman}

\usepackage{subfiles}
\title{\boldmath Ultraviolet Running Constraints on Low Mass Dark Sectors}
\author[a,b]{Aidan Reilly}
\author[a]{and Natalia Toro}
\affiliation[a]{SLAC National Accelerator Laboratory, 2575 Sand Hill Road, Menlo Park, CA 94025, USA}
\affiliation[b]{Physics Department, Stanford University, Stanford, CA 94305, USA}

\emailAdd{areilly8@stanford.edu, ntoro@slac.stanford.edu}

\abstract{We analyze the UV breakdown of Sub-GeV dark matter models that live in a new, dark U(1) sector. Many of these models include a scalar field, which is either the dark matter itself or a dark Higgs field that generates mass terms for the dark matter particle via spontaneous symmetry breaking. A quartic self coupling of this scalar field is generically allowed, and we show that its running is largely governed by the strength of the U(1) gauge field, $\alpha_D$. Furthermore, it consistently has a lower Landau pole than the gauge coupling. Link fields, which couple to both the dark sector and the Standard Model (SM), connect these Landau poles to constraints on SM charged particles. Current LHC constraints on link fields are compatible with $\alpha_D \lesssim 0.5 - 1$ for most of the mass range in most models, while smaller values,  $\alpha_D \lesssim  0.15$, are favored for Majorana DM.}

\graphicspath{{\subfix{images/}}}

\begin{document}
\maketitle
\flushbottom

\section{Introduction}
\label{intro}

The nature of dark matter (DM) is one of the largest outstanding problems in physics today. Over decades, more and more evidence for the existence of particle like DM has stacked up, but despite the same decades worth of searches, no DM candidate has been detected. This has motivated a broader search for DM particles of various origins. One of the more commonly discussed mechanisms for producing the correct relic abundance of DM involves thermalization with the Standard Model (SM) in the early universe, and subsequent freeze out of interactions as the universe cooled down. This idea of thermal freeze out is often associated with the WIMP miracle, wherein a  hypothetical supersymmetric particle that interacts with the SM only via the Weak force naturally produces the right abundance of DM (see \cite{Wimp} for a review). However, Weak scale particles are not the only viable candidates for thermalized DM, as minimal extensions to the SM can include Sub-GeV scale thermal relics as well. 

Sub-GeV thermal relic DM is compelling for a number of reasons. Such light particles are able to maintain a simple production mechanism while also evading current WIMP constraints \cite{LHC2,LHCsimp}. Furthermore, they require non-zero interactions with the SM such that they are potentially observable at direct detection (DD) and accelerator based experiments. In fact, there is growing interest in experiments proposed to search for such thermal relics, including low energy threshold DD experiments \cite{Essig:2022}, and high intensity and/or precision accelerator experiments based on production of DM and subsequent missing energy or scattering searches \cite{Ilten:2022, Krnjaic:2022}. Achieving correct relic abundance for these light particles typically requires a new GeV scale mediator\cite{Boehm:2002,Pospelov:2007,Arkani-Hamed:2008}. One of the simplest and most appealing scenarios involves a vector mediator that mixes with the SM hypercharge via kinetic mixing, which is widely used as a benchmark model \cite{CosmicVisions17, Fabbrichesi:2020, Graham:2021}. This interaction can be generated by loops of heavy particles (known as link fields), rendering it naturally small, and is common in extensions of the SM \cite{Holdom:1985, delAguila:1988}. 
After electroweak symmetry breaking, this vector, which we will refer to as a dark photon, couples to the electromagnetic current but with suppressed strength parameterized by $\epsilon$. In the simple case of a single link field, this interaction naturally produces $\epsilon \approx 10^{-1} - 10^{-3}$, while related mechanisms can generate smaller $\epsilon$ \cite{Gherghetta:2019, Cheung:2009, Schmitz:2009}.  
All of the DM candidates we consider here are part of this family of models, living in a new gauge sector, $U(1)_D$.

 The DM particle in such a gauge sector can be either a scalar or fermion, and while its relic abundance is indeed UV insensitive due to thermalization, the evolution of dark sector couplings to higher energy scales informs where in the parameter space to look,  and what is learned from experimental searches. Both gauge couplings and scalar quartic couplings grow in the UV. While previous papers have studied the effects of the gauge running \cite{DM}, we extend these analyses to include the running of a scalar quartic. Such a coupling can arise either directly from the DM itself, or from a dark Higgs sector, which is often included to restore symmetry to the low energy theory. As such, both scalar and fermionic Sub-GeV DM models are expected to have scalar quartic couplings. The poles of these quartic couplings tighten the constraints on $\alpha_D$ for Majorana DM, while for other models, competing effects lead to similar constraints as those usually assumed based on \cite{DM}.

\begin{figure}
    \centering
    \includegraphics[height=8cm, width=10cm]{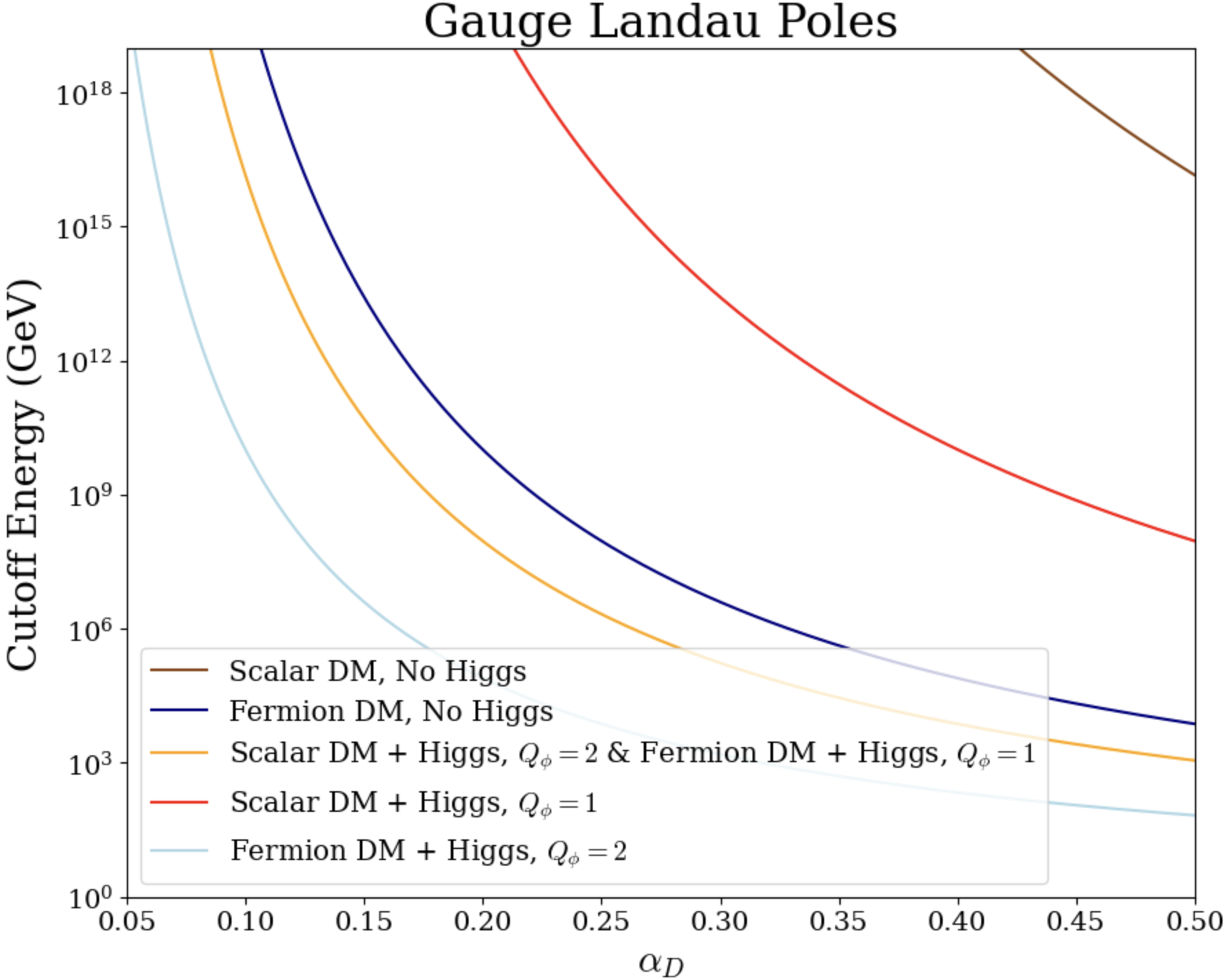} 
    \caption{Gauge Landau poles of various Higgsed scenarios with $m_\chi = 0.2$ GeV and $m_{A'} = 3m_\chi$.(Brown) Scalar DM with no Higgs, (Dark Blue) fermionic DM with no Higgs, (Red) scalar DM with Higgs of charge $Q_\phi = 1$, (Light Blue) fermionic DM with Higgs of charge $Q_\phi = 2$, (Orange) scalar DM with Higgs of charge $Q_\phi = 2$ and fermionic DM with Higgs of charge $Q_\phi = 1$.}
    \label{fig:alpha running}
\end{figure}

We start by reviewing the running of the gauge coupling studied in \cite{DM}, and the implication that it has for $\alpha_D = \frac{e_D^2}{4\pi}$. Throughout this paper we will take $\alpha_D$ to mean the value of the gauge coupling at the mass of the dark photon as is common in the literature, and we will explicitly write $\alpha_D(\mu)$ as a function of the energy scale $\mu$ when considering its running. By design, we often consider values of $\alpha_D$ very close to the perturbativity limit. Experiments are generally less sensitive to DM detection when $\alpha_D$ is large because generating the correct annihilation cross section for freeze out then demands a consequently smaller kinetic mixing paremeter. Thus, the strongest theoretically consistent coupling should be considered in order to probe all potential parameter space. A benchmark of $\alpha_D = 0.5$ is commonly used, motivated by the low gauge poles of more strongly coupled theories. Figure \ref{fig:alpha running} shows the gauge Landau pole for both fermionic and scalar kinetically mixed DM, with and without an additional Higgs field of different charges,  where the DM particle has unit charge $Q_\chi = 1$ ($g_D = Q_\chi e_D$). The light blue line in Figure \ref{fig:alpha running}, which corresponds to fermionic DM with a Higgs of charge $Q_\phi = 2$,  is discussed in \cite{DM}, where it was shown that  $\alpha_D$ hits a Landau Pole at energy $\mu^* \lesssim O(100 \text{ GeV})$ for $\alpha_D \gtrsim 0.5$.  This presents a problem because in the strong coupling regime the theory needs a UV completion, such as embedding the $U(1)_D$ group into a non-Abelian gauge theory. In order to generate a kinetic mixing parameter that is small, but not too small, \cite{DM} argues that a link field should enter at or below $\mu^*$, which is ruled out by collider experiments below the Weak scale \cite{DM_weak_Scale}. Thus, $\mu^* \sim O(100 \text{ GeV})$ can be viewed as constraining $\alpha_{D} \lesssim 0.5$ \cite{DM}, though the corresponding cutoffs (and maximum permissible values for $\alpha_D$) are higher for other DM spins and charges. Because Sub-GeV DM models motivate smaller kinetic mixing than considered in \cite{DM}, we find that link field masses somewhat above the symmetry breaking scale of the  non-Abelian gauge theory are valid, as discussed in Section \ref{link fields}. Nonetheless, the qualitative argument still holds: the need for light SM charged link fields near the scale of the gauge Landau pole, combined with non-observation of such new states at colliders, implies an upper bound on the dark gauge coupling that defines a ``worst case" target for dark sector searches.

We show in this paper that the running of a scalar quartic coupling, $\lambda$, implies similar, but more stringent bounds on $\alpha_D$. $\lambda$ typically encounters a Landau pole at parametrically lower energy scales than the gauge coupling, and it is driven predominantly by gauge contributions. As $ \lambda$ is not directly related to the DM effective field theory (EFT) operators relevant for freeze out or detection, most studies have chosen a small $\lambda$ under the assumption that we can therefore ignore its running. Sometimes, this idea is used as justification for a low dark Higgs mass $m_h$, which in turn implies certain phenomenology \cite{BelleII}. We show, however, that for sufficiently large $\alpha_D$, taking $\lambda$ small has a negligible effect on its running. The importance of $\alpha_D$ in the running of $\lambda$ was noted qualitatively in \cite{Darme:2018}, but we will make the relationship, as well as what it means for the theory, more precise.  

While completing the scalar quartic might in principle happen entirely in the dark sector, the simplest mechanisms involve adding charged particles that slow the running of $\lambda$. These particles will in turn lower the the pole of $\alpha_D$, which can be very constraining as just discussed. We will see that gauge Landau poles as low as a few 10's of GeV are consisent with current constraints. Demanding that the scalar quartic remain perturbative to the same energy scale constrains $\alpha_D$ to as low as $\lesssim 0.15$ in some models, and not below $0.5$ in others. If we instead consider the stronger criterion that the theory remain perturbative above collider scales, say to 100 TeV, then $\alpha_D$ becomes constrained to $\lesssim 0.05 - 0.2$, depending on the model.

The rest of this paper is organized as follows: in Section \ref{background} we introduce the models that we analyze including their completions by a dark Higgs mechanism and the full UV parameters used to calculate loop diagrams. In Section \ref{link fields} we analyze constraints based on generating the correct kinetic mixing parameter and how it relates to new SM charged particles. In Section \ref{poles} we discuss where UV completions are needed for all of the DM models presented in Section \ref{background} as a result of Landau poles. In Section \ref{Implications} we discuss one possible completion and implications for experiments. Finally, we discuss our results and make concluding remarks in Section \ref{conc}.

\section{Model Background}
\label{background}

In this paper we analyze models in which DM directly annihilates into SM particles through an intermediate dark photon mediator $A'$. This model is ubiquitous throughout the literature as a simple and relatively general mechanism for thermal DM \cite{CosmicVisions17, Fabbrichesi:2020, Graham:2021}. $A'$ is the massive gauge boson of a broken $U(1)_D$ symmetry. The dark photon kinetically mixes with the SM hypercharge, so that the relevant Lagrangian takes the form
\begin{equation}
    \mathcal{L} = \frac{\epsilon}{2\cos (\theta_W)}F'_{\mu\nu}B^{\mu\nu} + \frac{1}{2}m_{A'}A'_{\mu}A'^{\mu},
\end{equation}
where $\theta_W$ is the Weak mixing angle, $\epsilon$ is the kinetic mixing parameter, and $m_{A'}$ is the dark photon mass.  In the low-energy theory, $\epsilon$ is a free parameter, but $\epsilon \ll 1$ is often expected to be generated by loops of particles (link fields) charged under both hypercharge and $U(1)_D$ . In the limit that $m_{A'}\lesssim$ GeV, the dark photon dominantly mixes with the SM photon. After diagonalizing the kinetic and mass terms, $A'$ inherits an $\epsilon$-suppressed interaction with the electromagnetic current, $J_{EM}$, and retains an unsuppressed coupling to the $U(1)_D$ current, $J_D$. In the mass eigenstate basis, these take the form:

\begin{equation}
    -\mathcal{L} \supset A_\mu'(\epsilon e J_{EM}^\mu + g_DJ_D^\mu),
\end{equation}
where $g_D = \sqrt{4\pi \alpha_D}$. The form that $J_D$ takes depends on the specific model, generating different detection signatures. The relic abundance will be set by a DM DM $\rightarrow A' \rightarrow$ SM SM annihilation cross section: 
\begin{equation}
    (\sigma v)_{ann} \propto \frac{\epsilon^2\alpha_Dm_\chi^2}{m_{A'}^4}.
\end{equation}
All models carry this parametric dependence, however, $p$-wave annihilations carry an added velocity suppression relative to $s$-wave annihilations. We will focus our analyses on a representative model where $m_\chi = 0.2$ GeV and $m_{A'} = 3m_\chi$, but all Landau poles calculated scale linearly with DM mass. Changing the DM to dark photon mass ratio has a more complicated effect, but $m_{A'} = 3 m_\chi$ is a commonly chosen model as it it allows for on shell decays of $A'$ into DM and avoids resonances in the early universe \cite{ldmx}. For $m_\chi \lesssim m_A \lesssim 2m_\chi$, annihilation through an off shell A' competes with other (multi-body  or kinematically suppressed) processes 
\cite{DAgnolo:2020, Cline:2017, Fitzpatrick:2020, Fitzpatrick:2021}. On the other hand, as we increase the dark photon mass such that $m_A > 3 m_\chi$, DM becomes easier to see at accelerator searches, up to a maximum energy threshold \cite{Berlin:2020}.
 
\subsection{Dark Matter Models}
\label{unifications}

We discuss four specific models of vector portal DM that are viable thermal relics. In two of these models the DM is a scalar: Scalar Elastic Dark Matter (SEDM) and Scalar Inelastic Dark Matter(SIDM), while the in other two, the DM is a fermion: Pseudo-Dirac Dark Matter (PDDM) and Majorana Dark Matter (MDM). The four theories all have distinct phenomenology in the non-relativistic regime, but above the U(1) symmetry breaking scale, both scalar theories have the same particle content, as do both fermionic theories. 

\begin{center}
{\it 1. Scalar DM }
\end{center}
For both SEDM and SIDM we start with a complex scalar DM particle $\chi$ so that the relevant Lagrangian, including its coupling to the dark photon, is
\begin{equation}
\label{eq: scalar lagrangian}
\mathcal{L} \supset -ig_{D_0}(\chi^*\partial^\mu\chi - \chi\partial^\mu\chi^*)A'_\mu  - m_\chi^2 |\chi|^2 + \frac{1}{2}\mu_\chi^2(\chi ^2 + \chi^{*2}),
\end{equation}
which has a Dirac mass term $m_\chi|\chi|^2$, and a $U(1)_D$ symmetry breaking mass term $\mu_\chi$. However, this symmetry is already broken by the $A'$ mass, so $\mu_\chi$ should generally be considered. In the limit $\mu_\chi\rightarrow 0$, we recover the Lagrangian for a complex scalar and have SEDM. On the other hand, non-zero $\mu_\chi$ splits the masses of Re$[\chi]$ and Im$[\chi]$. Taking $\chi = \frac{1}{\sqrt{2}}(\chi_1 + i\chi_2)$, we find masses $m_{\chi_{1,2}} = \sqrt{m_\chi^2 \mp \mu_\chi^2}$ and a mass-off-diagonal current $J_D^\mu = \chi_2\partial^\mu\chi_1 - \chi_1\partial^\mu\chi_2$. In both cases, the dark gauge coupling of the effective current is the same as in the full theory, $g_{D_0}$. Left out of Eq. (\ref{eq: scalar lagrangian}) is the term $g_{D_0}^2A'^\mu A'_\mu|\chi|^2$, but it is only involved in subdominant processes during freeze out and detection.

\begin{center}
{\it 2. Fermionic DM }
\end{center}
To realize PDDM and MDM, we begin with the Lagrangian for a Dirac fermion that couples to $A'$:
\begin{equation}
\begin{split}
    \label{fermion lagrangian}
    \mathcal{L} &\supset g_{D_0}\bar{\chi}\gamma^\mu\chi A_\mu' - m\bar{\chi}\chi \\
    &=  -g_{D_0}(\eta^\dagger\bar{\sigma}^\mu A'_\mu\eta - \xi^\dagger\bar{\sigma}^\mu A'_\mu\xi) -m_D\eta\xi + h.c.,
\end{split}
\end{equation}
where in the second line we have split $\chi = \begin{pmatrix}
    \eta \\ \xi^\dagger
\end{pmatrix}$ into its two component Weyl spinors. Once again, we also allow $U(1)_D$ symmetry breaking mass terms so that the relevant Lagrangian becomes
\begin{equation}
\begin{split}
    \mathcal{L}\supset -m_D\eta\xi - \frac{1}{2}m_\eta\eta\eta - \frac{1}{2}m_\xi\xi\xi + h.c.
\end{split}
\label{original}
\end{equation}
 We can diagonalize the mass matrix $M = \begin{pmatrix}m_\eta & m_D \\ m_D & m_\xi\end{pmatrix}$ to find the mass eigenstates:

\begin{equation}
    \begin{split}
        \chi_l = \frac{i}{\sqrt{2}}\left(\eta - \left(\frac{m_\eta-m_\xi}{2m_D} + \sqrt{1 + \frac{(m_\eta-m_\xi)^2}{4m_D^2}}\right)\xi\right)
    \end{split}
\end{equation}
and 
\begin{equation}
    \chi_h = \frac{1}{\sqrt{2}}\left(\eta - \left(\frac{m_\eta-m_\xi}{2m_D} - \sqrt{1 + \frac{(m_\eta-m_\xi)^2}{4m_D^2}}\right)\xi\right)
\end{equation}
with 
\begin{equation}
    \label{eq: dm masses}
    m_{\chi_l,\chi_h} = \sqrt{m_D^2 + (m_\eta-m_\xi)^2/4} \mp \delta_\chi,
\end{equation} 
where $\delta_\chi = (m_\eta + m_\xi)/2$. Solving for $\eta$ and $\xi$ and substituting in Eq. (\ref{fermion lagrangian}), we find interactions with the dark photon that take the form:
\begin{equation}
\begin{split}
     \mathcal{L} &\supset -ig_{D_0} \frac{m_D}{\sqrt{m_D+(m_\eta-m_\xi)^2/4}}\bigg(\chi_l^\dagger\bar{\sigma}^\mu A'_\mu\chi_h - \chi_h^\dagger\bar{\sigma}^\mu A'_\mu\chi_l\bigg)\\
    &\quad -g_{D_0}\frac{(m_\eta-m_\xi)/2}{\sqrt{m_D^2+(m_\eta-m_\xi)^2/4}}\bigg(\chi_h^\dagger\bar{\sigma}^\mu A'_\mu\chi_h - \chi_l^\dagger\bar{\sigma}^\mu A'_\mu\chi_l\bigg).
\end{split}
\label{currents}
\end{equation}

If $m_\eta, m_\xi \ll m_D$, we end up with a small mass splitting and either suppressed or nonexistent elastic couplings, such that we have inelastic PDDM with a coupling to $A'$ given by the first line of Eq. (\ref{currents}). The associated PDDM current is therefore $J_D^\mu = i\bar{\chi_l}\gamma^\mu\chi_h$, with the effective gauge coupling $g_D=g_{D_0}$. 
On the other hand, if $m_\eta+m_\xi \gtrsim m_D$, then $\chi_l$ and $\chi_h$ have a sizeable mass splitting, and below $m_{\chi_h}$ we have just the one Majorana fermion, $\chi_l$. If we further have that $m_\eta \neq m_\xi$, $\chi_l$ gets a coupling to $A'$ given by the last term of Eq. (\ref{currents}). This generates the MDM current $J_D^\mu = \frac{1}{2}\bar{\chi_l}\gamma^\mu\gamma^5\chi_l$ with the effective gauge coupling $g_D = g_{D_0}(m_\eta-m_\xi)/\sqrt{4m_D^2+(m_\eta-m_\xi)^2}$, where we have added a $\frac{1}{2}$ to the current to account for identical particles in the Feynman rules. Thus, we have started with just one Dirac fermion with Majorana mass terms, and under two different limits, have arrived at either PDDM or MDM. Note that we could have just postulated a single Weyl fermion with axial coupling to $A'$ from the start, but it would be anomalous. Formulating the theory in the manner we have means that $\chi_h$ serves to cancel that anomaly. 

Finally, we should comment on two other limits of this theory. Taking $m_\eta+m_\xi \gtrsim m_D$ and $m_\eta = m_\xi$ leads to PDDM with large mass splitting. This model requires annihilation cross sections to increase exponentially with the size of the splitting, and added care when analyzing cosmological bounds, so we will not consider it \cite{ldmx}. We also do not take the elastic limit in the fermionic case. Elastic fermionic DM annihilates to SM particles via $s$-wave annihilations, which is ruled out by CMB constaints in the Sub-GeV range \cite{Planck:2015, Slatyer:2009}. The charge currents, effective couplings, and annihilation type of all four theories are summarized in Table \ref{table: eft ops} for reference.

\begin{center}
\begin{table}
\begin{tabular}
{|p{0.75cm}|p{3.0cm}|p{3.1cm}|p{2.75cm}|p{3.5cm}|}
 \hline
  & \multicolumn{2}{|c|}{{\bf Scalar Dark Matter}}  & \multicolumn{2}{|c|}{{\bf Fermionic Dark Matter}}  \\
 \cline{2-5}
 & \hfil {\bf Elastic} & \hfil {\bf Inelastic}&\hfil {\bf Pseudo-Dirac}&\hfil {\bf Majorana}\\
 \hline
 &&&&\\
 \hfil {\bf \it{J}}$_D^\mu$& $i(\chi^*\partial ^\mu\chi - \chi\partial ^\mu\chi^*)$   & \hfil $\chi_2 \partial ^\mu\chi_1 - \chi_1\partial ^\mu\chi_2$    &\hfil $ i\bar{\chi_l}\gamma^\mu\chi_h$& \hfil $\frac{1}{2}\bar{\chi_l}\gamma^\mu\gamma^5\chi_l$ \\
 &&&&\\
 \hline
 &&&&\\
 \hfil {\bf \it{g}}$_D$ & \hfil $g_{D_0}$ & \hfil $g_{D_0}$ & \hfil $g_{D_0}$ & \hfil $g_{D_0}\frac{(m_\eta-m_\xi)/2}{\sqrt{m_D^2+(m_\eta-m_\xi)^2/4}}$\\
 &&&&\\
 \hline
 &&&&\\
 \hfil {\bf $\sigma_{\text{ann}}$ } & \hfil p-wave & \hfil p-wave & \hfil s-wave & \hfil p-wave\\
 &&&&\\
 \hline
\end{tabular}
\caption{Important characterizations of the four DM models analyzed in this paper. The first row is the effective current, the second is the coupling strength of said current to the dark photon, and the third row is the the type of annihilation process relevant for freeze out.}
\label{table: eft ops}
\end{table}
\end{center}

\subsection{Dark Higgs Completion}
\label{higgs sector}
 Both the $A'$ mass and the mass terms $\mu_\chi$ and $m_{\eta,\xi}$ break the $U(1)_D$ symmetry. We can restore this symmetry in the UV by introducing a Higgs mechanism, wherein a new scalar field $\phi$ spontaneously breaks the $U(1)_D$ gauge symmetry, generating Majorana masses $m_\eta$ and $m_\xi$ in the fermionic case, and $\mu_\chi^2$ in the scalar case. It is worth mentioning that in all theories with $U(1)_D$ breaking masses for charged matter, we can reformulate the issue of gauge invariance as an issue of renormalizability under an appropriate gauge choice \cite{Craig}. Doing so imposes a clear upper bound on the breakdown of perturbative unitarity in the Higgsless theory. This makes explicit the need to introduce some symmetry restoring mechanism, an Abelian Higgs model being the simplest.  In general, the coupling $\chi^c\chi\phi^P$ is gauge invariant if $Q_\phi = \frac{2}{P}$. The only renormalizable couplings in 4 dimensions are $P \leq 2$ for scalar DM and $P=1$ for fermionic DM, and for any $P> 2 (1)$, the Higgs-scalar(fermion) coupling is irrelevant (i.e. has negative mass dimension). These higher P models are readily UV completed by introducing new particles at high energies whose charges permit renormalizable Higgs couplings. In fact, this gives a compelling  mechanism for generating the small mass terms required in PDDM (or SIDM). Throughout this paper we work only with dimension 4 and 5 couplings, though higher dimension couplings are possible, up to to a point, and are discussed in Appendix \ref{app: higher P theories}.  
 
 We take the Higgs sector of the form
\begin{equation}
\begin{split}
    \mathcal{L}_\phi \supset & |D_\mu\phi|^2+m_\phi^2|\phi|^2 - \frac{\lambda}{4}|\phi|^4,
\end{split}
\label{higgs}
\end{equation}
with $D_\mu = \partial_\mu - ig_\phi A_\mu'$, where $g_\phi = Q_\phi e_D$.  This potential for the Higgs field $V = -m^2|\phi|^2 + \frac{\lambda}{4}|\phi|^4$ gives rise to a vacuum expectation value (VEV) for $\phi$: $\langle \phi \rangle = \frac{v}{\sqrt{2}} =\sqrt{2m_\phi^2/\lambda}$. Expanding around this VEV as
\begin{equation}
    \phi = \frac{1}{\sqrt{2}}(v + h(x))e^{\frac{i\pi(x)}{v}},
\end{equation}
where $h$ and $\pi$ are both real scalar fields, we find a dark photon mass $m_{A'} = |g_\phi| v$. The Goldstone mode $\pi$ is eaten by the massive dark photon, while $h$ is a dark Higgs boson with mass $m_h = m_\phi \sqrt{2}$. Similarly, Higgs field couplings to DM of the form
\begin{equation}
    \mathcal{L}_{\phi \chi_s} \supset -y\phi^P\chi^2 + \text{h.c.}
\end{equation}
or 
\begin{equation}
    \mathcal{L}_{\phi \chi_f} \supset -\phi^P(\frac{1}{2}y_\eta\eta\eta + \frac{1}{2}y_\xi\xi^\dagger\xi^\dagger) + \text{h.c.}
\end{equation}
generate the symmetry breaking mass terms $\mu_\chi^2 = y\langle\phi\rangle^P$ or $m_{\eta,\xi}=y_{\eta,\xi}\langle\phi\rangle ^P$.

\section{Connecting Landau Poles to Collider Constraints}
\label{link fields}
In this section we hope to make clear the importance of understanding both gauge and scalar quartic Landau poles in these DM theories. The energy scale at which a parameter hits a Landau pole signals the need for a UV completion. One might expect that because UV completions of these theories can happen entirely in the dark sector, there would be no observable consequences. However, generation of kinetic mixing connects the UV completion scale with observable physics in the SM sector. The clearest UV completion for a gauge pole involves embedding the $U(1)_D$ into a non-Abelian gauge theory which is either asymptotically free or has an interacting fixed point \cite{Banks:1981nn, Litim:2014uca}. In non-Abelian theories, there is no renormalizable counterpart to the the kinetic mixing term that is invariant under the non-Abelian group symmetry. We can therefore generate kinetic mixing by integrating out loops of link fields, $L_i$, which are charged under both $U(1)_D$ and $U(1)_Y$, below the non-Abelian symmetry breaking scale. Otherwise, we can consider higher dimensional operators that connect $U(1)_Y$ to the dark sector above the non-Abelian symmetry breaking scale. Even in the latter case, the required dimensionality of the operator still leads us to consider a link field, only at higher masses. 

These link fields can be produced through Drell-Yan processes at colliders. Therefore, in order to avoid constraints on heavy stable charged particles, we can consider two possible decay signatures: (I) $L_i \rightarrow A'\ell_j$ where $\ell_j$ is a SM lepton, and (II) $L_i \rightarrow n W^{\pm}$ where $n$ is a new fermion with only dark charge, and $W^\pm$ is a W Boson. For more details about scenarios (I) and (II) we refer the reader to \cite{DM_weak_Scale}. Of particular interest to us is that we can use these decays to constrain the mass of $L_i$. Scenario (I) lends itself to di-lepton plus missing energy searches, while Scenario (II) can be probed with $WW$ plus missing energy searches. Analogous searches have been performed which set bounds on super partners in the $400$ to $700$ GeV range \cite{CMS:2020,ATLAS:2018}. We expect similar exclusions for link fields, so we will take $400$ GeV as a limit compatible with missing energy searches. Worth noting is that in scenario (II) we might avoid these constraints by setting the mass of $n$ very close to $m_{L_i}$, but it would put us in an extremely narrow region of parameter space. Furthermore, if we had a scalar link field instead of a fermion, we could replace the SM leptons in scenario (I) with a SM Higgs or gauge boson, and the fermion $n$ in scenario (II) with a scalar, and achieve qualitatively similar bounds.

To illustrate how this constraint restricts the energy scale at which the theory is embedded it into a non-Abelian completion, we will consider an example gauge coupling strength of $\alpha_D = 0.5$, a single link field of mass $M_L$, and SSB of $SU(N)_D$ to $U(1)_D$ at a scale $\mu^*$. If $M_L < \mu^*$, then integrating out loops of this link field in the Abelian theory leads to the kinetic mixing parameter
\begin{equation}
    \label{km param}
    \frac{\epsilon}{\cos\theta_W} = \frac{g_Yg_D}{16\pi^2}\ln(\frac{\mu^{*2}}{M_L^2}) \approx 5\times 10^{-3},
\end{equation}
 where $g_Y$ is the hypercharge coupling constant, and we have taken $\ln(\frac{\mu^{*2}}{M_L^2}) \sim \mathcal{O}(1)$. On the other hand, if $M_L > \mu^*$, it cannot directly generate kinetic mixing, which is forbidden by the gauge symmetry. It can, however, generate the operator $\text{Tr}[(\tilde{\phi}/\Lambda) \tilde{F}^{\mu\nu}]B_{\mu\nu}$, where $\tilde{F}$ is a dark gauge field in the adjoint of $SU(N)_D$ and $\tilde{\phi}$ is a scalar field charged under $\tilde{F}$ which will spontaneously break the $SU(N)_D$ symmetry into $U(1)_D$ \cite{Arkani-Hamed:2008}, with
 \begin{equation}
     \frac{1}{\Lambda} \approx \frac{g_Yg_D\tilde{y}}{16\pi^2}\frac{1}{M_L},
 \end{equation}
 where $\tilde{y}$ is the yukawa coupling between the link field and $\tilde{\phi}$, including factors of N associated with the size of the symmetry group. When we then take $\tilde{\phi}$ to its VEV, $\langle\tilde{\phi}\rangle \simeq \mu^*$, it induces the kinetic mixing
 \begin{equation}
    \label{eq: km param 2}
    \frac{\epsilon}{\cos\theta_W} \approx \frac{g_Yg_D\tilde{y}}{16\pi^2}\frac{\mu^*}{M_L} \approx 5\times 10^{-3} \frac{\mu^*}{M_L},
 \end{equation}
 for $\tilde{y} \sim \mathcal{O}(1)$.  Notice that Eq. (\ref{eq: km param 2}) always allows for a heavier link field mass than Eq. (\ref{km param}), so bounds on $M_L$ when kinetic mixing happens above $\mu^*$ imply the same or stronger bounds when kinetic mixing happens below $\mu^*$. A link field mass slightly less than one order of magnitude above $\mu^*$ generates $\epsilon \approx 7 \times 10^{-4}$, which, for our benchmark parameters, sets the correct thermal relic abundance for scalar DM. Requiring the link field mass to be $\gtrsim 400$ GeV therefore implies $\mu^* \gtrsim 40$ GeV. To generalize this relationship, we can define a link field scale
 \begin{equation}
 \label{eq: muL}
     \mu_L(\alpha_D, \epsilon) = \frac{g_Yg_D\mu^* \cos{\theta_W}}{16\pi^2 \epsilon},
 \end{equation}
which sets the link field mass as $M_L = \mu_L / \tilde{y}$. For subsequent analyses we set $\tilde{y} =1 $, but it can vary depending on the specifics of the non-Abelian theory.  Furthermore, in all models discussed here, the required $\epsilon$ for relic abundance scales nearly linearly with $m_\chi$, and so too do the Landau poles. Thus, even though the poles get lower with decreasing DM mass, the link field constraints weaken accordingly so that the $\alpha_D$ bound is roughly mass independent. Important to note is that accelerator based link field searches also creates a potential pathway towards discovery, or exploration of a signal detected elsewhere. Each value of $\alpha_D$ implies is a maximum allowed mass of the link field, which is a particle well suited for discovery at colliders. 
 
\begin{figure}
    \centering
    \includegraphics[height=8cm, width=10cm]{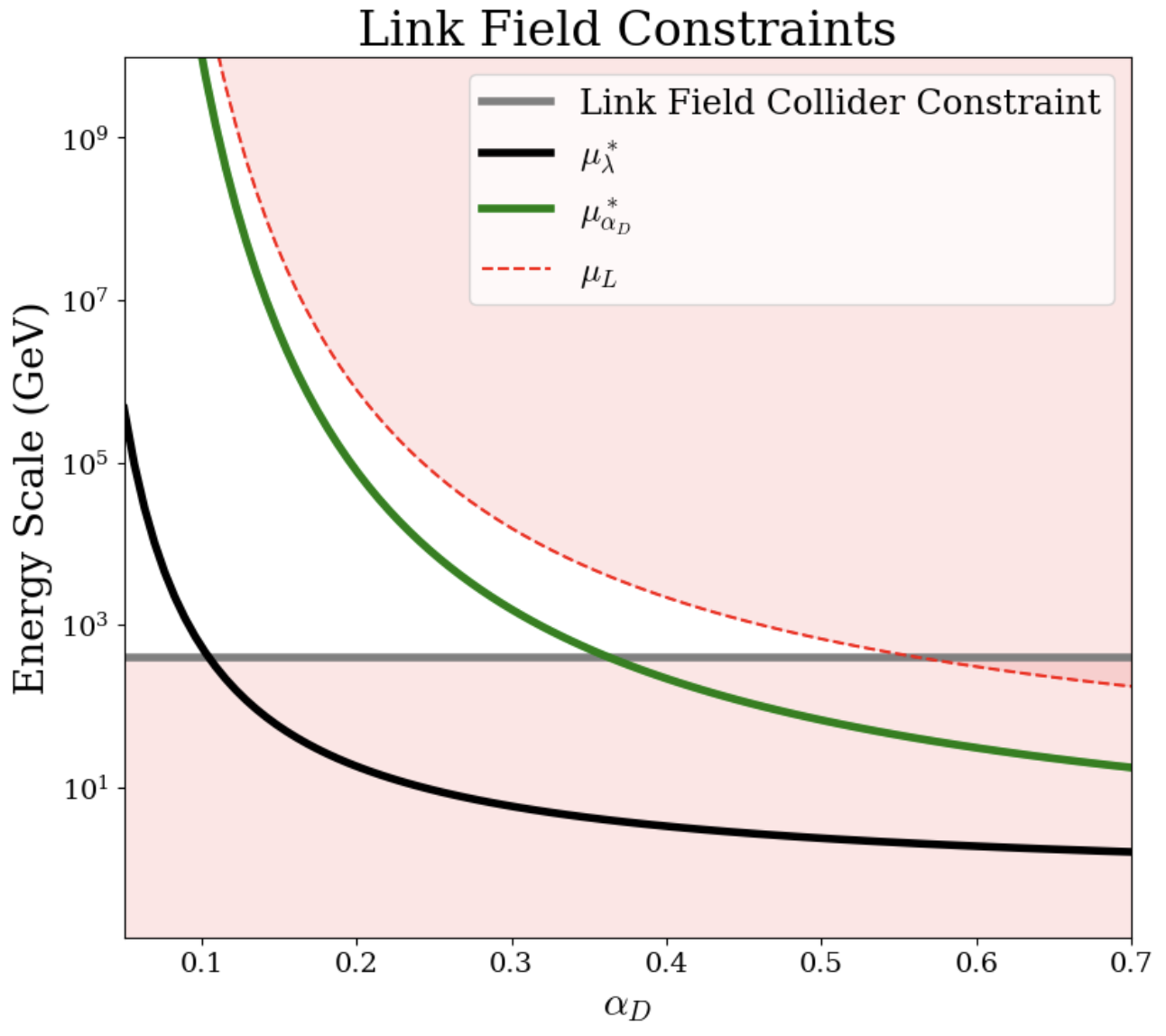} 
    \caption{Red shaded regions are excluded based on link field mass requirements. The black line shows how the quartic landau pole ($\mu^*_\lambda)$ is parametrically below the gauge pole ($\mu^*_{\alpha_{D}}$). Raising the $\lambda$-pole tends to lower the gauge pole, and consequently, the maximum link field mass.}
    \label{fig: link field energies}
\end{figure}

We can then turn our attention to the scalar quartic coupling, which  hits a Landau pole at lower energies than the gauge coupling. This demands its own UV completion, which can come in two forms. We can either add new particles which slow the running of $\lambda$, or we can make the Higgs a composite particle. Both of these situations inevitably involve new charged matter, which will lead to a lower required non-Abelian symmetry breaking scale. Thus, low $\lambda$ poles will effectively tighten the link field constraints as well. Figure \ref{fig: link field energies} shows this idea diagramatically, and we examine these completions more closely in Section \ref{novel completions}.

\section{Energy Scales for Loss of Perturbativity}
\label{poles}
 In this section we show at what energy scales each DM model discussed in Section \ref{background} hits a Landau pole, and what the consequences are for allowed parameter space. We wish to know what ranges of couplings are viable, as well as what the Landau poles can tell us about mass hierarchies in the dark sector. The former is addressed by the discussion in Section \ref{link fields}, and to address the latter, in Section \ref{higgs regions} we will classify regions of parameter space into one of three categories, which have distinct phenomenology. In Section \ref{inelastic}, we analyze the UV growth of inelastic, and elastic scalar, DM models, where DM-Higgs interactions are small enough to have negligible impact on $\beta$-functions. The simpler $\beta$-function for $\lambda$ in these models allows us to analytically solve for the running couplings $e_D(\mu)$ and $\lambda(\mu)$ (where $e_D=\sqrt{4\pi\alpha_D}$). Then, in Section \ref{majorana} we analyze the UV growth for MDM which has a large mass splitting. In this case, $\beta_\lambda$ includes terms proportional to $y_{\eta,\xi}$, which also run according to  $\beta_{y_{\eta,\xi}}$. Accordingly, we solve for $e_D(\mu)$ and $y_{\eta,\xi}(\mu)$ analytically, but rely on numerics to find the strong coupling limit of $\lambda$. Finally, in Section \ref{stueckelberg} we consider a SEDM theory where the dark photon has a Stueckelberg mass such that a Higgs field is never introduced. Even in this case, there is still an allowed quartic coupling of the DM particle to itself, $\lambda_\chi$, whose running we analyze. In all cases, we work to 1-loop order.
 
 \subsection{Regions of Different Higgs Phenomenology}
 \label{higgs regions}
\begin{figure}
    \centering
    \includegraphics[height=6cm, width=14cm]{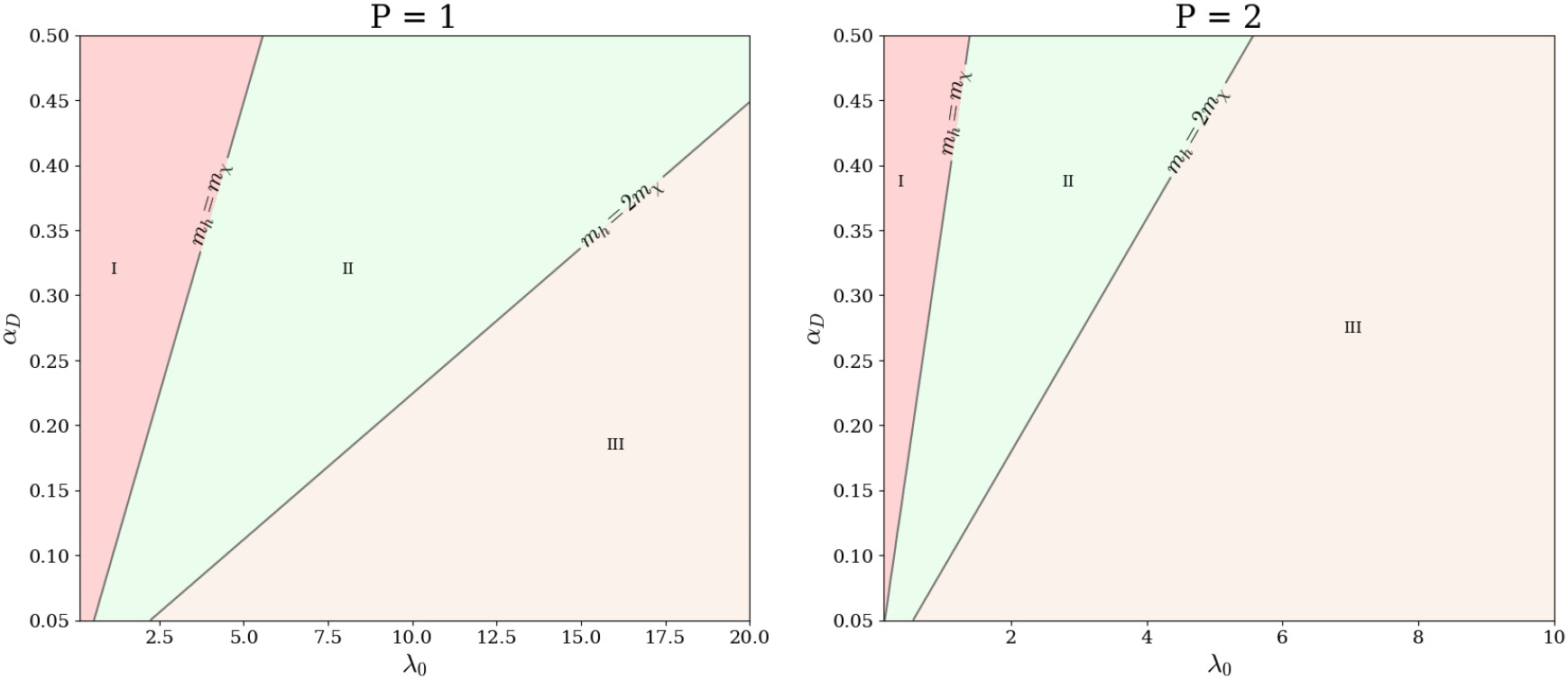} 
    \caption{Mass of the dark Higgs boson when $m_\chi = 0.2$ GeV and $m_{A'} = 3m_\chi$. On the left is a Higgs theory with $P = 1$ ($Q_\phi = 2$), and on the right a theory with $P = 2$ ($Q_\phi = 1$). Region I corresponds to parameter space where $m_h < m_\chi$, Region II corresponds to parameter space where $m_\chi < m_h < 2m_\chi$, and Region III corresponds to parameter space where $2m_\chi < m_h$.}
    \label{fig:mass space}
\end{figure}

 Different regions of the $\alpha_D - \lambda_0$ plane correspond to different masses of the dark Higgs boson for a given DM and dark photon mass, where $\lambda_0 = \lambda(m_h)$. Figure \ref{fig:mass space} breaks up this plane into three regions, which each have distinct phenomenology. Understanding which regions are or are not disfavored by low Landau poles will consequently inform what DM signatures are best motivated. To that end, we will overlay Landau poles with the regions of Figure \ref{fig:mass space} in the following sections so that the parameter space can be readily understood. Note that while Figure \ref{fig:mass space} only includes $P=1$ and $P=2$ models, later plots will include $P=3$. We can use the relationships $\lambda = 2Q_\phi^2e_D^2 (\frac{m_h}{m_{A'}})^2 \text{ and }  m_{A'}=3m_\chi$ to define the following regions:
 \begin{itemize}
     \item \underline{Region I (Unviable)}: $m_h \leq m_\chi$

 For a dark Higgs lighter than the DM particle, DM can efficiently annihilate into dark Higgses, which then decay only to SM particles. This depletes the DM abundance, making it cosmologically unviable. Once the dark Higgs is heavier than the DM particle, freeze out is largely set by the annihilation channel $\chi\chi \rightarrow A' \rightarrow \text{SM } \text{SM}$, and can produce the correct relic abundance. There is some parameter space where $m_h \gtrsim m_\chi$ such that thermal effects continue to over annihilate DM via $\chi\chi \rightarrow h h$ \cite{DAgnolo:2015}. These `forbidden' annihilation channels mean that just having $m_h > m_\chi$ is not a strict limit for safety, but the exact calculation for how much heavier $m_h$ must be than $m_\chi$ is beyond the scope of this paper. 
 
     \item \underline{Region II (Visible)}: $m_\chi < m_h < 2m_\chi$

 When $m_\chi < m_h < 2m_\chi$, dark Higgses will have bright decay signatures at collider experiments as their only decay products are SM particles. 

    \item \underline{Region III (Invisible)}: $ 2m_\chi < m_h$

 Once the dark Higgs is twice the DM mass it will readily decay invisibly back into DM.
\end{itemize}
We will therefore refer to I as the unviable region, II as the visible dark Higgs region, and III as the invisible dark Higgs region. Of interest is whether the allowed parameter space favors region II or region III, in which case light DM physics implies certain dark Higgs phenomenology, or whether the two are comparably open. For example, when $m_h$ is the lightest dark sector particle, decays to SM particles via Higgs mixing can be searched for at Belle II \cite{BelleII}. If Region II were to found to be the only viable parameter space, a lack of these signals would rule out the theory. On the other hand, if Region III remains open, one would have to consider an analogous missing energy search.  A more in depth discussion of these regions and potential signatures can be found in Appendix \ref{app: pheno}.

 \subsection{Negligible Majorana-like Masses}
 \label{inelastic}
\begin{figure}
    \centering
    \includegraphics[height=8cm, width=10cm]{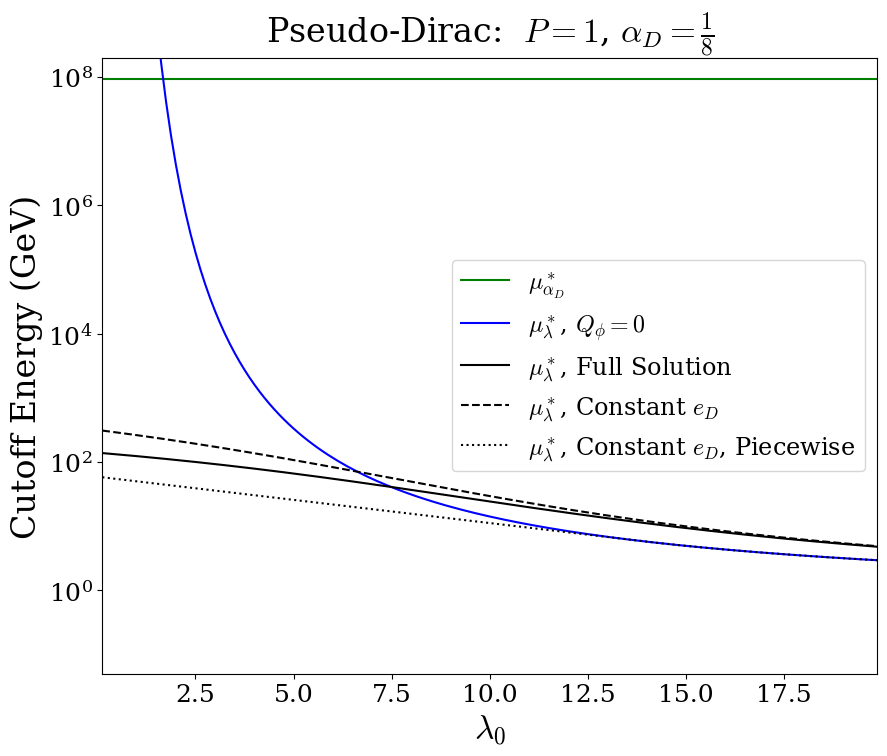} 
    \caption{Landau Poles of PDDM with $m_\chi = 0.2$ GeV and $m_{A'} = 3m_\chi$ at one particular $\alpha_D$ point. The blue line indicates a $\phi^4$ theory with no gauge contribution. The black dashed line indicates solving Eq.(\ref{eq:Be pddm}) under the approximation that $e_D(\mu)=e_{0}$. The solid black line indicates an exact solution to Eq.(\ref{eq:Be pddm}).  The black dotted line indicates solving Eq.(\ref{eq:Be pddm}) under the assumption of $e_D(\mu) = e_{0}$ and taking only the largest contribution to running at each step. The green line refers to the gauge pole.}
    \label{fig:l running}
\end{figure}

 We begin our analysis of $\beta$-functions with the simplest Higgs model to analyze: those where Majorana-like masses are small (SEDM, SIDM, PDDM), and so do not have yukawa couplings which contribute to the running of $e_D(\mu)$ or $\lambda(\mu)$. We choose to work in the unbroken phase of the theories described in Section \ref{background} for clarity, as $\beta$-functions are the same regardless of what VEV we expand around. In this regime, we simply have a gauge theory with one complex scalar (the Higgs field), one Dirac fermion or second complex scalar (the DM particle), and a quartic scalar potential. One can find a standard computation of the one loop $\beta$-functions $\beta_{e_D}$ and $\beta_\lambda$ for this theory from \cite{Srednicki}:
\begin{equation}
    \begin{split}
        \beta_{e_D} = \frac{1}{12\pi^2}(\sum_{\psi}Q_\psi^2 + \frac{1}{4}\sum_{\Phi}Q_\Phi^2)e_D^3 = ce_D^3 ,
    \end{split}
    \label{eq:Be pddm}
\end{equation}
\begin{equation}
    \begin{split}
        \beta_\lambda = \frac{1}{16\pi^2}(5\lambda^2 - 12\lambda(Q_\phi e_D)^2 + 24(Q_\phi e_D)^4), 
    \end{split}
    \label{eq:Bl pddm}
\end{equation}
where $\sum_{\psi}$ is a sum over all fermions in the theory and $\sum_{\Phi}$ is a sum over all scalars. We have written $c = \frac{1}{12\pi^2}(\sum_{\psi}Q_\psi^2 + \frac{1}{4}\sum_{\Phi}Q_\Phi^2)$ to highlight the fact that this parameter depends on the number of charged particles (and their corresponding charges) in the theory. For example, for fermionic DM without a Higgs field, $c=\frac{1}{12\pi^2}$, but with the Higgs field where $P=1$, $c=\frac{1}{6\pi^2}$, which has a sizeable effect on the location of the gauge Landau pole. Eq. (\ref{eq:Be pddm}) can be readily solved by direct integration, yielding the solution:
\begin{equation}
    \begin{split}
        e_D(t)^2 = \frac{e_0^2}{1-e_0^22c(t-t_0)}, 
    \end{split}
    \label{eq:e(u) pddm}
\end{equation}
where $t = \ln(\mu)$ is just the log energy scale. Eq. (\ref{eq:Bl pddm}), on the other hand, is more complicated because it has contributions from both $\lambda$ and $e_D$. Important to note, is that although Eq. (\ref{eq:Bl pddm}) has a negative term, by completing the square one can show that it is always positive. We derive an analytic solution in Appendix \ref{app: Bl} by mapping (\ref{eq:Bl pddm}) to the Riccati equation, which we use in the subsequent analysis. To get an idea for its general behavior, however, a much simpler piece-wise picture suffices. With the full solution at hand, it becomes clear that $e_D$ does not run very much before $\lambda$ hits a pole, so we can make the simplifying assumption  $e_D \simeq e_0$ in (\ref{eq:Bl pddm}). Furthermore, because $\beta_\lambda$ is always positive, either the first or third term must always dominate. Thus, we might see the IR to UV evolution as follows: at low scales, $\beta_\lambda$ is driven by a large constant term $\kappa = \frac{24(Q_\phi e_0)^4}{16\pi^2}$. Eventually $\lambda$ grows enough that the first term takes over so that $\beta_\lambda$ is driven by $b_{xx}\lambda^2 = \frac{5\lambda^2}{16\pi^2}$. The Landau poles, and hence the breakdown of perturbative validity, are then given by the values of $\mu^* = e^{t^*}$ that send $e_D,\lambda \rightarrow \infty$. The Landau pole of $e_D$ is an unsurprising one, though we should note, once again, that by including the charged Higgs field, we significantly increase the running and lower the location of the gauge Landau pole:
\begin{equation}
\label{eq:gauge pole}
    t^*_e = \frac{1}{2ce_0^2} + t_0.
\end{equation}
Important to see, however, is the dependence of $\lambda$ on both $\lambda_0$ and on $e_D$. The Landau pole for $\lambda$ is strictly below the Landau pole for $e_D$, regardless of how low we take $\lambda_0$. In the piece-wise picture, this is because $\lambda$ first runs as $\lambda(t) = \kappa(t-t_0) + \lambda_0$ up until $\lambda(t') = \lambda ' = \sqrt{\frac{\kappa}{b_{xx}}}$ at $t' = (\lambda' - \lambda_0)/\kappa + t_0$. The pole is then given by:
\begin{equation}
    \label{eq: piecewise pole}
    t^*_\lambda = t' + \frac{1}{\lambda' b_{xx}} = t_0 - \lambda_0/\kappa + \frac{2}{\sqrt{\kappa b_{xx}}}.
\end{equation}
Upon careful analysis of this equation, one finds that $t_\lambda^* < t_e^*$ regardless of $\lambda_0$ because $\frac{2}{\sqrt{\kappa b_{xx}}} < \frac{1}{2ce_0^2}$, as long as $Q_\phi \gtrsim 0.7$. This is obviously not satisfied for $P \geq 3$, in which case one needs to examine the full solution to see that $t_\lambda^* < t_e^*$. We plot the poles of this piece-wise function, alongside the poles where we only make the constant $e_D$ approximation, and the poles of full analytic solution in Figure \ref{fig:l running}, which demonstrates that this approximate picture is accurate up to a factor of a few. We also include the Landau poles of a theory where we have taken $Q_\phi \rightarrow 0$ to show the extreme effect that the gauge contribution has to $\lambda$ running.

\begin{figure}
    \centering
    \includegraphics[height=13cm, width=15cm]{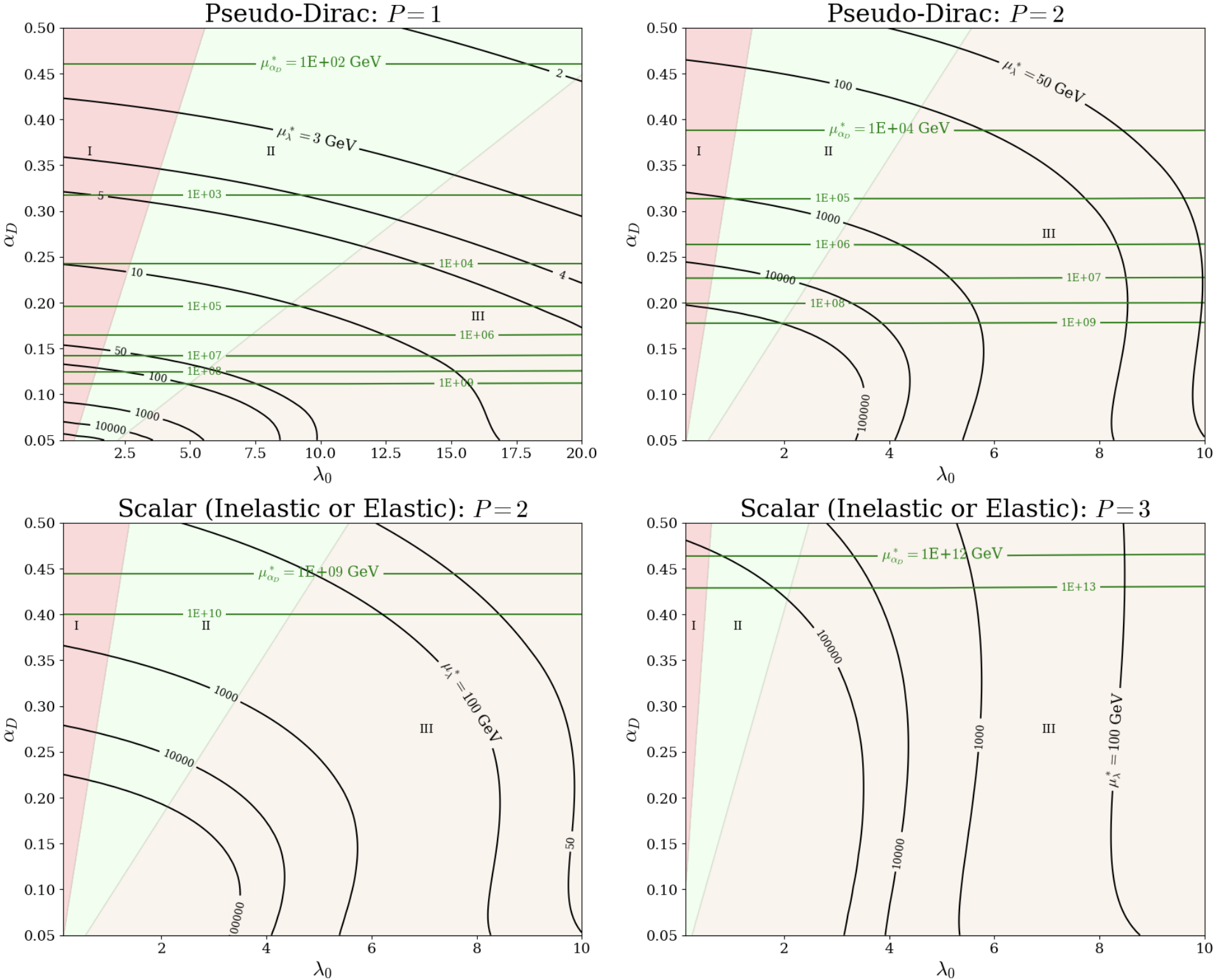} 
    \caption{Landau poles of PDDM, SIDM, and SEDM theories in GeV, with $m_\chi = 0.2$ GeV and $m_{A'} = 3m_\chi$. Green lines are gauge poles and black lines are poles of $\lambda$. Regions I-III correspond to the same regions as Figure \ref{fig:mass space}. Note that for SEDM the Higgs might not couple to DM at all, and thus the phenomenology of region III may not apply.} 
    \label{fig:inelastic poles}
\end{figure}

In Figure \ref{fig:inelastic poles}, the $\lambda$ poles as well as the gauge poles are plotted as contours on the $\alpha_D - \lambda_0$ plane. The top panels show fermionic PDDM, and the bottom panels show scalar DM that can correspond to either SEDM or SIDM, though the shaded regions (Higgs phenomenology) assume a Higgs-DM coupling, which can be zero for SEDM. For PPDM $P=1$ theories, $\alpha_D \gtrsim 0.15$ results in Landau poles $\leq 100$ GeV, while for PPDM $P=2$ theories, $\alpha_D \gtrsim 0.5$ results in the same. This makes the benefit of lowering the Higgs charge abundantly clear, while also motivating novel completions for both. Scalar $P=2$ theories have marginally higher cutoffs than PDDM $P=2$ theories, while scalar $P=3$ theories have significantly higher poles for low $\lambda_0$. Because regions I and II get more narrow with increasing $P$, breakdown of perturbativity does not particularly motivate having a Higgs in region II as opposed to region III for any theory. It is also worth commenting that in many regions of parameter space, $\lambda$ is large enough that higher order loop effects could potentially play an appreciable role in the running, but we leave this to future work. 

 \subsection{Majorana Dark Matter}
 \label{majorana}
\begin{figure}
    \centering
    \includegraphics[height=8cm, width=10cm]{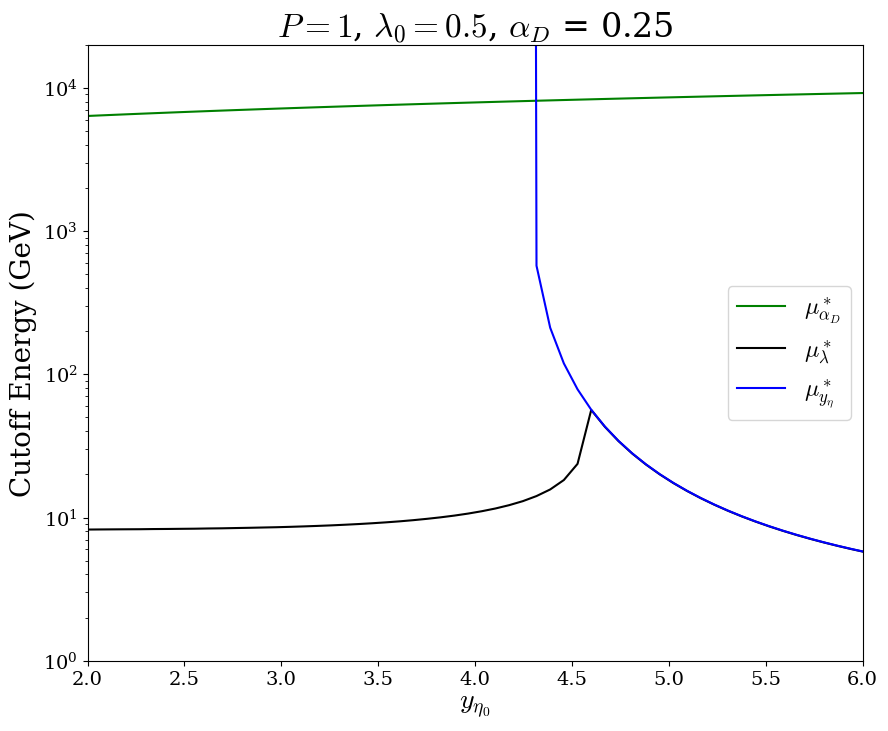} 
    \caption{Landau Poles for MDM with $m_\chi = 0.2$ GeV and $m_{A'} = 3m_\chi$. The green line is the gauge pole, the black line is the $\lambda$ pole, and the blue line is the Yukawa pole. We ignore contributions from loops of light state DM particles as they have negligible effect. Given that $\beta_\lambda$ depends on $y_\eta$, $\lambda$ becomes unstable at large $y_{\eta_0}$, allowing us to match up cutoffs of $\lambda$ and $y$ for $y_{\eta_0}\gtrsim 4.6$. }
    \label{fig:mdm opt}
\end{figure}

To extend the preceding analysis to the MDM model, we must include the Higgs-DM coupling in the renormalization group evolution. Furthermore, the only allowed charge of the Higgs is twice that of the DM particle ($P=1$) for MDM. Higher dimension couplings would suppress the Majorana mass term. To calculate the exact one loop contributions we use the results for general gauge theories in \cite{RGE}. Unlike PDDM where $y_{\eta,\xi}$ are fixed to be approximately $0$, we have more freedom in MDM. For a given $m_\chi$ and $m_{A'}$, there is still a wide range of values for $y_{\eta,\xi}$ and $m_D$ that satisfy Eq. (\ref{eq: dm masses}), which differ, however, in the heavy partner state $m_{\chi_h}$ and in the effective gauge coupling.  To simplify the problem, we restrict $y_\xi$ to $0$. This is a reasonable approximation because an unsuppressed Majorana coupling requires $m_\eta - m_\xi \gtrsim m_D$ (Table \ref{table: eft ops}), and appreciable Yukawa contributions to $\beta_\lambda$ require $m_\eta$ or $m_\xi \gg m_\chi$, which together with Eq. (\ref{eq: dm masses}) imply that one Yukawa coupling must be significantly greater than the other. The $\beta$-functions for all couplings in the model are then as follows
\begin{equation}
\label{eq:Be mdm}
    \begin{split}
        \beta_{e_D} = \frac{1}{12\pi^2}(\sum_{\psi}Q_\psi^2 + \frac{1}{4}\sum_{\Phi}Q_\Phi^2)e^3 = ce^3,
    \end{split}
\end{equation}
\begin{equation}
\label{eq:Bl mdm}
    \beta_\lambda = \frac{1}{16\pi^2}\left[5\lambda^2 - 4y_\eta^4 + 2y_\eta^2\lambda - 12g_\phi^2\lambda + 24g_\phi^4\right],
 \end{equation}
 \begin{equation}
 \label{eq:By simplified}
        \beta_{y_\eta} = \frac{1}{16\pi^2 }(\frac{3}{2}y_\eta^3 - 6g_\eta^2y_\eta).
\end{equation}
$\beta_{e_D}$ is exactly the same as in PDDM, alluding to the fact that, in the UV, these models are equivalent. Therefore, Eq. (\ref{eq:Be mdm}) emits the same solution and the same Landau pole.  We also solve Eq. (\ref{eq:By simplified}) analytically in Appendix \ref{app: By}, but Eq.~(\ref{eq:Bl mdm}) must be integrated numerically. Our algorithm finds the energy at which the gradient is above a large threshold, which signals an approach to infinity. The resulting Landau pole depends on $y_{\eta_0}$, in addition to $e_{D}$ and $\lambda_0$. The dependence is nontrivial: as $y_{\eta_0}$ increases, the $-y_{\eta}^4$ term in $\beta_\lambda$ starts to dominate and slow the running. However, at the same time, increasing $y_{\eta_0}$ increases the running of $y_\eta$ itself. Figure \ref{fig:mdm opt} shows an example of this process for representative values of $\alpha_D$ and $\lambda_0$. We scan over a range of values $y_{\eta_0} \in (1,7)$, choosing at each point in the $\alpha_D - \lambda_0$ plane the value of $y_{\eta_0}$ that maximizes the minimum of all poles in the theory. As the cutoff for $\lambda$ is always lowest, this amounts to maximizing the $\lambda$ cutoff. In our optimization, we ignore contributions to running from loops of only light state DM particles. Light state contributions are heavily suppressed so this approximation has no qualitative effect. 

\begin{figure}
    \centering
    \includegraphics[height=8cm, width=10cm]{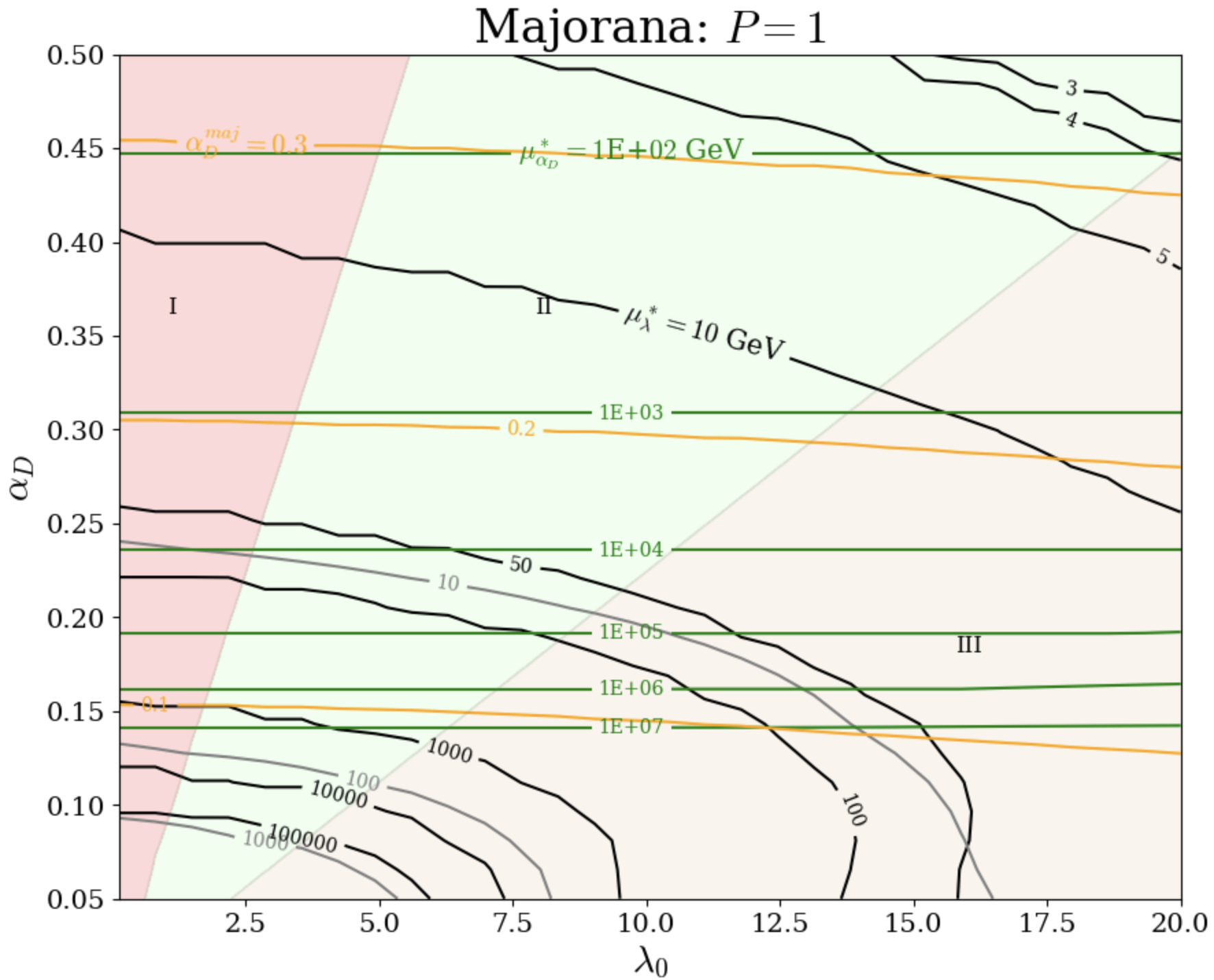} 
    \caption{Landau Poles of MDM in GeV with $m_\chi = 0.2$ GeV and $m_{A'} = 3m_\chi$. Green lines are gauge poles, black lines are optimized poles of $\lambda$, gray lines are non-optimized poles of $\lambda$, and orange lines correspond to effective $\alpha_D$ couplings in the low energy MDM model. Regions I-III correspond to the same regions as Figure \ref{fig:mass space}. The approximation is made here that DM contributions to running do not start until the mass of the heavier state.}
    \label{fig:mdm poles}
\end{figure}

In Figure \ref{fig:mdm poles} we plot these poles as contours in the $\alpha_D - \lambda_0$ plane. We also show a few contours that correspond to $y_{\eta_0} \sim 1$ to highlight the effect of optimizing $y_{\eta_0}$. Important to recall is also the fact that, while in PDDM $\alpha_D$ of the UV theory corresponds to $\alpha_D^{\text{eff}}$ of the low energy DM model, in MDM there is a suppression in the effective coupling compared to the UV parameter $\alpha_D$, demonstrated by the orange contours. In MDM, remaining perturbative to $\gtrsim 100$ GeV requires $\alpha_D^{maj} \lesssim 0.15$. This is roughly the same as PDDM $P=1$, because the benefit of negative Yukawa contributions to $\lambda$ running is almost exactly cancelled by the suppression in $\alpha_D^{maj}$ relative to $\alpha_{D}$.

 \subsection{Stueckelberg Scalar Dark Matter}
 \label{stueckelberg}
\begin{figure}
    \centering
    \includegraphics[height=8cm, width=10cm]{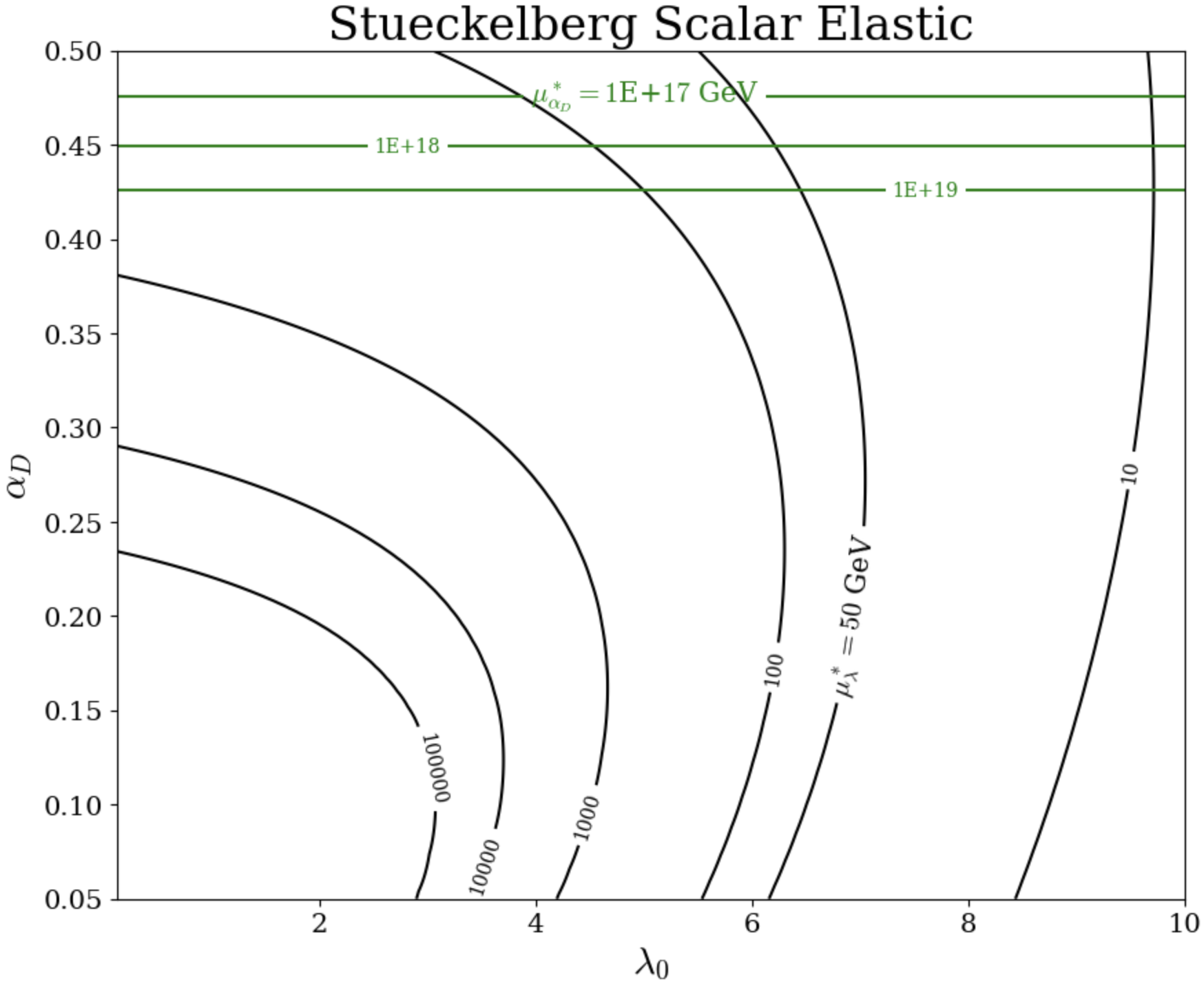} 
    \caption{Landau poles of Stueckelberg SEDM in GeV with $m_\chi = 0.2$ GeV and $m_{A'} = 3m_\chi$. Green lines are gauge poles and black lines are poles of $\lambda_\chi$.}
    \label{fig:stueckelberg poles}
\end{figure}

The last situation we need to analyze is the Higgsless SEDM model, otherwise known as a Stueckelberg theory. In SEDM we can take $\mu_\chi \rightarrow 0$, so we need not introduce any Higgs field. Instead, we can employ the `Stueckelberg trick,' which one can think of as a Higgsed theory with $m_\phi$ taken to $ \infty$ \cite{Stueckelberg:1938}. While this theory has no Higgs quartic coupling to worry about, the scalar DM field also admits a dimensionless quartic self-coupling consistent with all symmetries. The relevant Lagrangian in this case becomes
 \begin{equation}
    \mathcal{L} \supset \frac{1}{2}|D_\mu\chi|^2 - m_\chi^2\chi^2 - \frac{1}{4}{F'}_{\mu\nu}^2 + \frac{1}{2}m_{A'}{A'}^2 - \frac{1}{4}\lambda_\chi|\chi|^4.
 \end{equation}
 Luckily, the $\beta$-function for $e_D$ is once again left unchanged, just with fewer charged particles running in loops and thus a lower $c$ in Eq.~(\ref{eq:Be pddm}).  Similarly, the running for $\lambda_x$ will take the same form as that of $\lambda$ in Eq.~(\ref{eq:Bl pddm}). Thus, both couplings admit the same solutions with appropriate charge counting. In Figure \ref{fig:stueckelberg poles} we plot contours of the Landau poles in this theory. Although somewhat less severe than in the other models simply because there are fewer charged particles running in loops, Landau poles below a few hundred GeV are still expected for $\alpha_D=0.5$, even when the low energy quartic coupling is taken to vanish. Further, the quartic Landau pole is again the leading constraint, not the gauge Landau pole. It is also worth noting, once again, the potential for 2-loop effects to play a role.
 
 Interestingly, the size of $\alpha_{D}$ and $\lambda_\chi$ are directly related to the DM self interaction cross section, on which there are limits set by astrophysical observations. Typical observations suggests that the self interaction cross section should not exceed $\approx O(0.1 - 1 \text{ cm}^{2} \text{ g}^{-1}$) \cite{Markevitch:2003, Eckert:2022}. The DM  self interaction cross section is smaller than $0.1 \text{ cm}^{2} \text{ g}^{-1}$ everywhere on this plot, and regions with high Landau poles have substantially lower cross sections. We should mention that the Higgsed versions of SEDM and SIDM possess the same DM self interaction quartics, but in those cases their effects are less than that of the Higgs quartic. Since we can set the initial value of $\lambda_\chi < \lambda$, and it otherwise runs in the same way, the Higgs qaurtic will always hit a Landau pole first. 

\section{Implications of UV Completing $\lambda$}
\label{Implications}
 We have seen that gauge Landau poles motivate a significant structural change to the dark sector, such as embedding the $U(1)_D$ into a non-Abelian group, and that $\lambda$ hits a pole at parametrically lower energy scales. The $\lambda$ Landau pole could have a similarly drastic completion, for example, making the dark Higgs composite. However, the SM suggests a more incremental change in the form of adding fermions with strong couplings to the Higgs. The top quark coupling results in a large negative contribution which slows the SM Higgs quartic running, and in fact, we took advantage of this same general feature in Section \ref{majorana} when we optimized the Higgs-DM interaction strength. Adding new $U(1)_D$ charged matter lets us generalize this completion to models where the DM itself has small or higher-dimension Yukawa couplings. We discuss this completion in Section \ref{novel completions}, and in Section \ref{pheno} we analyze the implications for experiment. 

\subsection{UV Completing $\lambda$}
\label{novel completions}
In all of the models discussed so far, the gauge pole is above the $\lambda$ pole, sometimes significantly so. One might try to take advantage of this by introducing new fermions which couple to the scalar field in question and act to slow down the running of $\lambda$. In the presence of a new Dirac fermion (and in the absence of significant DM-Higgs couplings), $\beta_\lambda$ takes the form
\begin{equation}
    \beta_\lambda = \frac{1}{16\pi^2}\left[5\lambda^2 -8y^4 + 4y^2\lambda - 12g_\phi^2\lambda + 24g_\phi^4\right],
\end{equation}
while the $\beta$-function for this new Yukawa coupling takes the form
\begin{figure}[htb!]
    \centering
    \includegraphics[height=8cm, width=10cm]{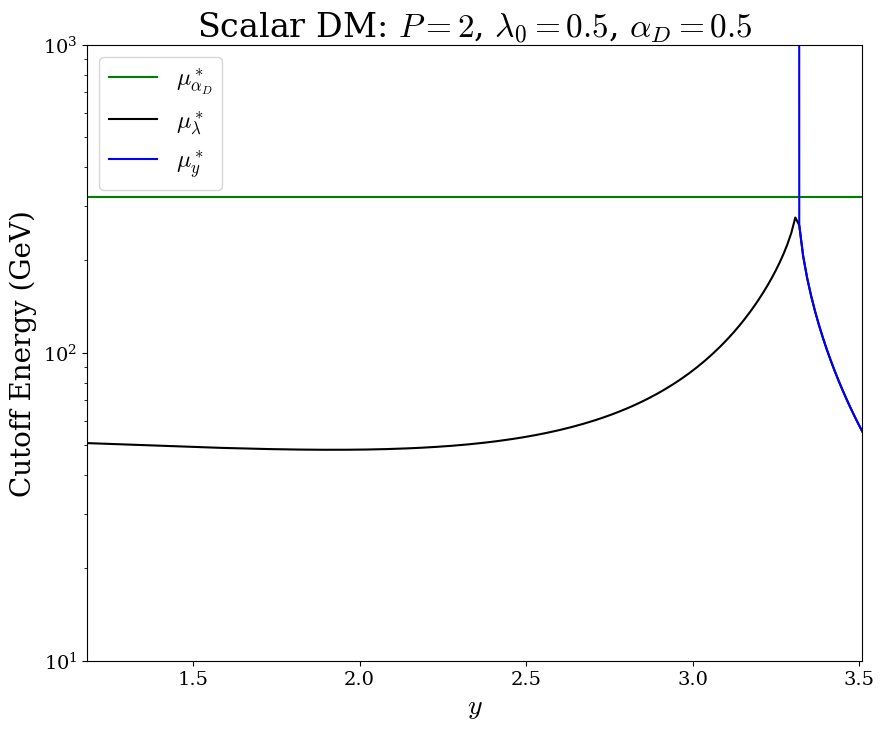} 
    \caption{Landau Poles for Scalar DM with $m_\chi = 0.2$ GeV, $m_{A'} = 3m_\chi$ and a new fermion with a Dirac mass set to make it the heaviest particle in the theory as we vary the Yukawa coupling. The green line is the gauge pole, the black line is the $\lambda$ pole, and the blue line is the Yukawa pole of the new fermion.}
    \label{fig:new fermion poles}
\end{figure}
\begin{equation}
    \beta_y = \frac{1}{16\pi^2}\left[2y^3 - 6g_f^2y\right],
\end{equation}
where $g_f = Q_f e_D(\mu)$ is the gaugle coupling of the new fermion. Clearly if $y$ is large enough, it can stop the running of $\lambda$, but if it gets too large, it will hit its own Landau pole sooner. In Figure \ref{fig:new fermion poles} we plot the Landau poles for scalar DM as a function of the Yukawa coupling for a new fermion which has a Dirac mass only big enough to make it the largest particle in the theory. Note that we ignore the mass splitting of this new particle as it has negligible consequence on the running in the parameter space of interest. We are able to achieve some incremental improvements, but a full completion really will require multiple new fermions that enter near the non-Abelian symmetry breaking scale. 
We note an approximate fixed point, where we add multiple fermions with weaker couplings such that neither $y$ nor $\lambda$ run appreciably. Letting $n$ be the number of new fermions with identical couplings to the Higgs, we can cancel the dominant gauge contribution to $\beta_\lambda$ if
\begin{equation}
    ny^4 \simeq 3g_\phi^4,
\end{equation}
and $\beta_y \approx 0$ if 
\begin{equation}
    y^2 \simeq 3g_f^2 = \frac{3}{4}g_\phi^2.
\end{equation}
Satisfying both requires $n \simeq \frac{16}{3} \sim 5$ new fermions. This seems promising, but we have thus far ignored the gauge running. The addition of 5 new charged fermions means that $\alpha_D(\mu)$ will run into a Landau pole at very low energies. For example, if we set all 5 of these fermions to have mass $m_\psi = m_{A'}$ in a P=1 PDDM model where $\lambda_0 = 2.5$ and $\alpha_D = 0.25$, the gauge pole moves from $\mu_{\alpha_D}^* \simeq 10^4$ GeV  to  $\mu_{\alpha_D}^* \simeq 12$ GeV. The $\lambda$ Landau pole in such a theory without these new fermions would be $\mu_\lambda^* \simeq 10$ GeV, so we clearly do not buy ourselves anything by adding these fermions at this scale. While the quantitative relationship between the addition of new fermions and the gauge pole highly depends on the specifics of the theory, this example highlights the general effect.

In order to avoid siginificantly lowering the gauge pole, we can embed the $U(1)_D$ into a non-Abelian group near the scale where these fermions enter. Doing so leaves room for $\lambda$ to be completed perturbatively, though we do not pursue a specific completion in the non-Abelian sector. Interestingly, this connects the non-Abelian symmetry breaking scale, and consequently the mass of the link fields, to the scale of the $\lambda$ landau pole. We should keep this in mind as we analyze experimental results in the next section.

\subsection{Constraints on Minimal Models}
\label{pheno}  
Simple UV completions motivate thinking about the $\lambda$ Landau pole as governing the scale of link field masses. Even more complicated completions, such as making the Higgs composite, require new charged matter that likely motivates the same connection. We might therefore view a $\lambda$ pole, combined with experimental limits on link fields, as directly constraining $\alpha_D$ the same way that a gauge pole does. With that in mind, we can analyze what current link field constraints imply about the allowed size of $\alpha_D$ and the impact of these constraints on low-energy experiments' parameter space. As discussed in Section \ref{intro}, we will be primarily interested with missing energy experiments. In principle, lowering $\alpha_D$ has a similar effect on the sensitivity of beam dump  plus subsequent scattering searches, but it will be weaker. Thermal targets move up in $\epsilon$ space proportionally to $\alpha_D$, while sensitivity of missing energy searches does not change. Beam dump scattering searches, on the other hand, will decrease sensitivity proportionally to $\sqrt{\alpha_D}$. We will therefore show only the more severely affected case.

\begin{figure}[htbp]
    \centering
    \includegraphics[height=18cm, width=15cm]{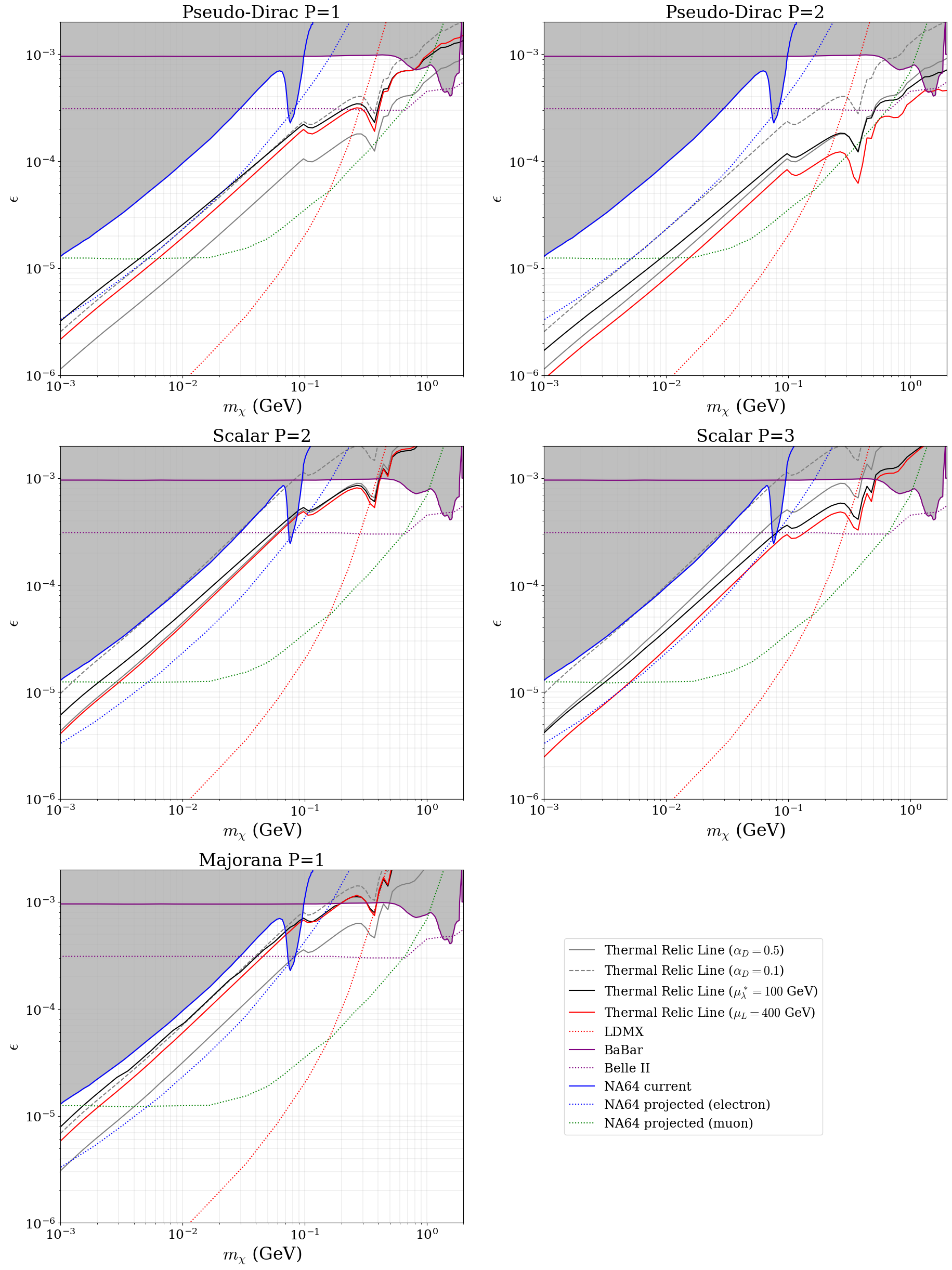} 
    \caption{Thermal targets for representative dark matter candidates. The black, gray, and red curves in each panel represent the parameter space for which the abundance of $\chi$ is in agreement with the observed dark matter energy density, where $\lambda_0$ is set to the lowest possible value such that $m_h \geq m_\chi$. The shaded region above the purple curve is excluded by the BaBar $\gamma^+$ missing energy search \cite{Izaguirre:2013, Essig:2013}. The dashed purple curve is the projected sensitivity of a $\gamma^+$ missing energy search at Belle II using $20 \text{ fb}^{-1}$ \cite{CosmicVisions17, Belle-II:2022cgf} 
    . The shaded region above the solid blue curve is excluded by the NA64 missing energy search for DM with $\alpha_D = 0.1$ \cite{NA64}. The dashed blue and green lines are the projected capabilities for an NA64 electron and muon beam respectively. The red dashed curve is the LDMX projection for a 10\% radiation length tungsten target and an 8 GeV beam presented in \cite{Hearty:2017} , which was scaled up to $10^{16}$ EOT relative to a background study with 4 $\times 10^{14}$ EOT \cite{ldmx}.}
    \label{fig:exp reach}
\end{figure}

In Figure \ref{fig:exp reach} we plot contour lines in the $\epsilon - m_\chi$ plane that correspond to the correct thermal relic abundance of DM for each model with a fixed $\alpha_D$.  Since the annihilation rate of $\chi\chi \rightarrow \text{SM SM}$ is what dictates these lines, they are linearly proportionally to $\alpha_D$, and thus get shifted accordingly. On the same plot, we show lines of current and projected experimental reach for a few beam dump experiments whose capabilities are independent of $\alpha_D$. Note that current NA64 limits are actually $\alpha_D$ dependent, but only when the incoming $e^+e^-$ are resonant with $A'$. Furthermore, for all $\alpha_D\lesssim 0.1$ this resonance peak asymptotes to the same value. As such, we still choose to include these limits with a set $\alpha_D$ of 0.1. On these plots we also include thermal relic lines that correspond to Landau poles at a fixed energy. These lines take on varying $\alpha_D$ values, but correspond to the maximum $\alpha_D$ should we demand the theory not break down below a certain scale. Similarly, the plots feature lines of fixed $\mu_L$ as defined in Eq. (\ref{eq: muL}), where $\mu^* = \mu^*_\lambda$. We should make clear that here we have not included any new fermions in our analyses.

One can notice that even for cutoff energies just barely allowed by collider constraints, fixed $\alpha_D$ of $0.5$ may be too large (i.e. $\mu_L \leq 400$ GeV). However, current collider constraints pose no problem to those theories with Landau poles even as low as 100 GeV. Table \ref{table: cutoffs} shows the values of $\alpha_D$ which correspond to $\mu_L = 400$ GeV and 100 TeV for each model. There is a small range in $\alpha_D$ between $m_\chi = 1$ meV and $m_\chi = 2$ GeV due to the deviation in linear scaling of $\epsilon$ above 0.1 GeV, with a greater variation for larger values of $\alpha_D$. The first row of this table, corresponding to the red lines in Figure \ref{fig:exp reach}, gives a realistic estimate for the maximum value of $\alpha_D$ in each model. Should one of these DM models be detected with an $\alpha_D$ in the range of the second row, a collider with energy $>$ 100 TeV would be required to verify its link field physics. Note that although Scalar DM and PDDM theories with the same dark Higgs charge have similar $\lambda$ poles, Scalar DM requires a larger kinetic mixing parameter to be cosmologically viable, so the bounds on $\alpha_D$ are stronger. It merits calling out once again that higher P theories, when allowed, are the preferred models from the point of view of  avoiding Landau poles. This does mean, however, that for light masses, theories with high $\mu^*_\lambda$ might actually breakdown earlier due to the dimension 5 Higgs - DM coupling. This can be easily fixed, however, with the addition of a neutral fermion when $P=2$, and a fermion with charge $1/3$ when $P=3$, thus not affecting the gauge pole much.

\begin{center}
\begin{table}
\begin{tabular}
{|p{2.75cm}|p{1.9cm}|p{1.9cm}|p{1.9cm}|p{1.9cm}|p{1.9cm}|}
 \cline{1-6}
\center  & {\bf PDDM P=1}& {\bf PDDM P=2} & {\bf Scalar DM P=2} &{\bf Scalar DM P=3}& \bf {MDM P=1}\\
 \hline
 &&&&&\\
 \hfil $\mathbf{\mu_L}$ {\bf = 400 GeV} & 0.14 - 0.19 & 0.77 - 2.0 & 0.56 - 0.67 & 1.5 - 2.0  &  0.14 - 0.17 \\
 &&&&&\\
 \hline 
 &&&&&\\
 \hfil $\mathbf{\mu_L}$ {\bf = 100 TeV} & .059 - .065 & 0.24 - 0.27 & 0.24 - 0.25 & 0.55 - 0.59  &  .065 - .065 \\
 &&&&&\\
 \hline
\end{tabular}
\caption{Values of $\alpha_D$ for each model which correspond to a given link field mass scale ($\mu_L$). The lower limit is for $m_\chi = 1$ meV and the upper limit is for $m_\chi = 2$ GeV, with $m_{A'}=3m_\chi$. We also demand here that $\mu^*_\lambda \gtrsim 3m_{A'}$ for the theory to be reasonable.}
\label{table: cutoffs}
\end{table}
\end{center}

\section{Conclusion}
\label{conc}

In this paper, we have analyzed the non-perturbative regimes of various Sub-GeV vector portal DM candidates. The importance of avoiding a gauge Landau pole is well known in these models, as values for $\alpha_D$ are taken to be as general as possible, including those that put the theory near a gauge pole. Here, we extended the analyses of Landau poles to other parameters, with particular focus on the dark Higgs quartic self coupling parameter $\lambda$ present in these models. We have shown that, for large $\alpha_D$ commonly considered in the literature, the running of $\lambda(\mu)$ is strongly governed by the size of $\alpha_D$, and not very sensitive to the low energy Higgs quartic. Furthermore, this running results in Landau poles that are always below the gauge Landau pole. We connect the location of these poles to collider constraints on SM charged link fields, effectively restricting the low energy  parameter $\alpha_D$. These analyses make it clear that for PDDM, SEDM, and SIDM, theories where the dark Higgs charge is equal to or less than the DM charge are preferrred, from the perspective of having higher Landau poles. They also contrast the idea that a light dark Higgs is theoretically favored by purtubativity arguments. We then discussed one possible method of slowing down the $\lambda$ running via the addition of new fermions. In order for this to constitute a full UV completion, the theory must also be embedded into a non-Abelian gauge sector near the $\lambda$ pole. Demanding that the minimal theory remain perturbative up to energies where this is possible improves the theoretical reach of some DM searches. Finally, the possibility of testing the running of $\alpha_D$ by performing experiments at different energies was outlined in \cite{DM}. It could be interesting to think about analogous dark Higgs signatures. However, while the running of $\lambda$ is more dramatic over the same range of energies, it would require looking at multi-dark Higgs production processes which is experimentally challenging. We leave the exploration of such signatures to future work.

\acknowledgments

We thank Maxim Pospelov and Philip Schuster for stimulating discussions that partially motivated this work. The authors were supported by the U.S. Department of
Energy under contract number DE-AC02-76SF00515 while at SLAC, and Aidan Reilly was
supported by the NSF GRFP under grant DGE-2146755.

\bibliographystyle{JHEP}
\bibliography{main}

\appendix

\section{Solving $\beta_\lambda$ in the Negligible Majorana Mass Limit}
\label{app: Bl}
In this appendix we analytically solve the following two differential equations
\begin{equation}
    \begin{split}
        \beta_{e_D} = \frac{1}{12\pi^2}(\sum_{\psi}Q_\psi^2 + \frac{1}{4}\sum_{\Phi}Q_\Phi^2)e^3 = ce^3, 
    \end{split}
\end{equation}
\begin{equation}
    \begin{split}
        \beta_\lambda = \frac{1}{16\pi^2}(5\lambda^2 - 12\lambda(Q_\phi e)^2 + 24(Q_\phi e)^4) ,
    \end{split}
\end{equation}
where of course $\beta_x = \frac{dx}{d \ln{\mu}} = \frac{dx}{dt}$ where $\mu$ is an energy scale.  $\beta_{e_D}$ is easily solved by separation of variables to find
\begin{equation}
    \begin{split}
        e_D(t)^2 = \frac{e_0^2}{1-e_0^22c(t-t_0)} ,
    \end{split}
    \label{eq:e(u) pddm}
\end{equation}
which hits a pole at 
\begin{equation}
    t^*_e = \frac{1}{2ce_0^2} + t_0.
\end{equation}
$\beta_\lambda$ is a little more complicated, and we start by defining a variable $X = (\lambda - \frac{6}{5}Q_\phi^2 e_D^2)$ which we can differentiate to find
\begin{equation}
\begin{split}
\label{eq: BX}
\beta_X = b_{xx}X^2 + b_{xg}e_D(t)^4,
\end{split}
\end{equation}
where $b_{xx} = \frac{5}{16\pi^2}$, $b_{xg} = \frac{84Q_\phi^4}{5\cdot 16\pi^2} - \frac{12cQ_\phi^4}{5}$. Now we can note that any differential equation of the form $y' = q_0(t) + q_1(t)y + q_2(t)y^2$ which has $q_0,q_2 \neq 0$ is known as a Riccati equation, and can be transformed from a non-linear, first order differential equation into a linear, second order differential equation of the form:
\begin{equation}
    u'' - Ru' + Su = 0,
\end{equation}
 via the substitutions $R = q_1 + \frac{q_2'}{q_2}$, $S=q_2q_0$, and $y = -\frac{u'}{q_2 u }$. We can see that, for Eq. (\ref{eq: BX}), $R=0$ and $S = b_{xx}b_{xg}e_D(t)^4$. From here we simply plug $u''(t) = b_{xx}b_{xg}e_D^4(t) u(t)$ into Mathematica DSolve \cite{Dsolve} to arrive at the solution:
\begin{equation}
    u(t) = f(t)^{\frac{1-z}{2}}(c_1 + c_2f(t)^z)
\end{equation}
where $z = \frac{\sqrt{b_{gg}^2 - 4b_{xg}b_{xx}}}{b_{gg}}$ and $f(t) = (-1+g_0tb_{gg})$, $b_{gg} = 2c$, $g_0 = e_{0}Q_\phi$, and $c_{1,2}$ are arbitrary coefficients that we will fix with boundary conditions. We can then recover $X(t)$ by computing $-\frac{u'(t)}{q_2 u}$ to find
\begin{equation}
    \begin{split}
        X(t) = \frac{e^2(t)b_{gg}}{b_{xx}}\left[(\frac{1-z}{2}) + \frac{z}{\frac{c_1}{c_2f(t)^z}+1}\right].
    \end{split}
\end{equation}
We see that $c_1$ and $c_2$ only show up as a multiplicative constant $\frac{c_1}{c_2}$ which we will call $C$. This makes sense given that our originally problem was a first order differential equation. We can then solve for $C$ in terms of the initial condition $X_0$ to find
\begin{equation}
    C = e^{\pm\pi|z|}\left[\frac{K-z}{K+Z}\right],
\end{equation}
where $K = 2\frac{X_0b_{xx}}{g_0b_{gg}} - 1$ and $\pm \pi$ is chosen to always give a real solution. Finally, we can obtain an expression for $\lambda(t)$:
\begin{equation}
    \lambda(t) = X(t) + \frac{6}{5}Q_\phi^2 e_D^2 = \frac{e_D^2(t)b_{gg}}{b_{xx}}\left[(\frac{1-z}{2}) + \frac{z}{\frac{C}{f(t)^z}+1} + \frac{6 Q_\phi^2b_{xx}}{5b_{gg}}\right],
\end{equation}
which hits a pole at
\begin{equation}
  t^*_\lambda = \frac{1 + (-C)^\frac{1}{z}}{b_{gg}e_0^2} + t_0
    = t_e^* + \frac{(-C)^\frac{1}{z}}{b_{gg}e_0^2},
\label{eq:lpole pddm}
\end{equation}
and while it may not be immediately apparent, $\frac{(-C)^{\frac{1}{z}}}{b_{gg}e_0^2}$ is always negative so that $t_\lambda^* < t_e^*$.

\section{Solving $\beta_{y_\eta}$ in the $y_\xi = 0$ Limit}
\label{app: By}
In this appendix we analytically solve the beta function for the relevant Yukawa coupling in MDM 
\begin{equation}
        \beta_{y_\eta} = \frac{1}{16\pi^2 }(\frac{3}{2}y_\eta^3 - 6Q_\eta^2e_D^2(t)y_\eta).
\end{equation}
Recalling that $e_D(t)$ is a known function of $t$, we can recognize $\beta_{y_\eta}$ as a Bernoulli differential equation which takes the canonical form $y' + P(t)y = Q(t) y^n$ with 
\begin{equation}
    \begin{split}
        P(t) &= \frac{6Q_f^2}{16\pi^2}e_D^2(t), \\
        Q(t) &= \frac{3}{32\pi^2},\\
        n &= 3.
    \end{split}
\end{equation}
We can thus use the typical tricks for solving Bernoulli's equation, starting with the substitution of $u=y^{1-n}$. Plugging that in lets us re-write our equation as
\begin{equation}
    \frac{du}{dt} - (n-1)P(t)u = -(n-1)Q(t),
\end{equation}
which gives a first order, linear, inhomogenous differential equation of the form
\begin{equation}
    u' + p(t)u = q(t),
\end{equation}
noting that $p(t) = -(n-1)P(t)$ and $q(t) = -(n-1)Q(t)$. We can solve this using an integrating factor
\begin{equation}
\begin{split}
    I(t) &= e^{\int p(t)dt} = |1 -e_0^22c(t-t_{0e})|^{\frac{(n-1)6Q_f^2}{2c16\pi^2}},
\end{split}
\end{equation}
where $t_{0e}$ is where the gauge running starts. We can then relate $I(t)$ to $u(t)$ via the equation
\begin{equation}
\begin{split}
    I(t)u(t) &= \int I(t)q(t) dt .
\end{split}
\end{equation}
Integrating and solving for $u$ gives us
\begin{equation}
u = \frac{3}{16\pi^2}\left[\left( \frac{1-b(t-t_{0e})}{(a+1)b} \right) + C_u(1-b(t-t_{0e}))^{-a}\right],
\end{equation}
where $a = \frac{(n-1)6Q_f^2}{1c16\pi^2}$, $b= 2ce_0^2$, $t_{0e}$ is the where the gauge running starts, and $C_u$ is a constant set by the boundary condition $u_0 = y_{0\eta}^{-2}$:
\begin{equation}
C_u = (1 -b(t_{0y}-t_{0e}))^{a}\left[\frac{16\pi^2}{3}y_0^{-2} - \left( \frac{1-b(t_{0y}-t_{0e})}{(a+1)b}\right)\right].
\end{equation}
And thus, we have the equation
\begin{equation}
y_\eta(t) = \left[\frac{3}{16\pi^2}\left[\left( \frac{1-b(t-t_{0e})}{(a+1)b} \right) + C_u(1-b(t-t_{0e}))^{-a}\right]\right]^{-1/2},
\end{equation}
which has the a pole at
\begin{equation}
    t_y^* = \frac{1-\left[C_u(a+1)b\right]^{\frac{1}{a+1}}}{b} + t_{0e}.
\end{equation}
It is worth noting that when $y_\eta$ running starts before the gauge running, the $\beta$-function and consequently $y_\eta (\mu)$ would take a different form until the gauge contributions kicked in. However, throughout the analyses done here, this is never the case.
\section{Higher P Theories}
\label{app: higher P theories}
\begin{figure}
    \centering
    \includegraphics[height=6cm, width=15cm]{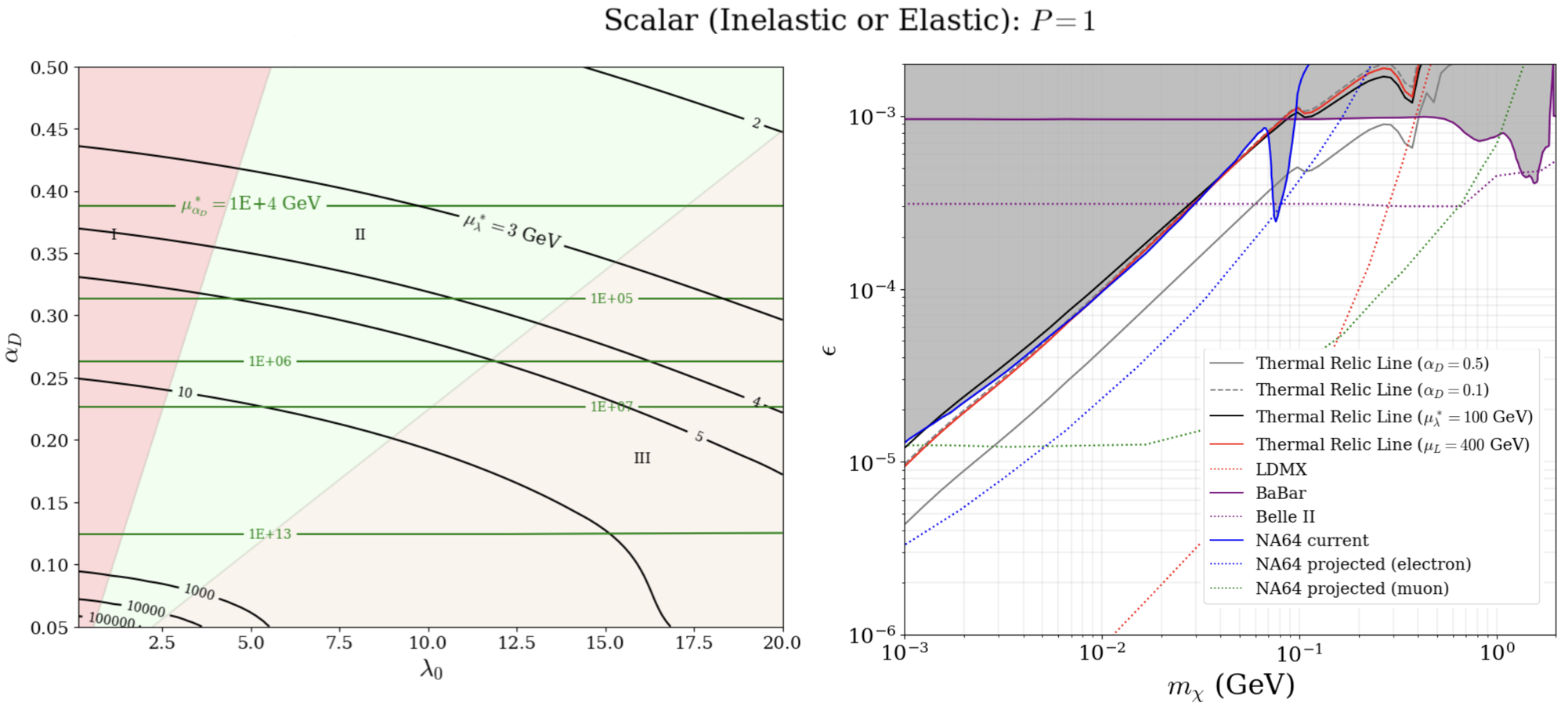} 
    \caption{ (Left) Landau Poles of Scalar DM in GeV with $m_\chi = 0.2$ GeV and $m_{A'} = 3m_\chi$. Green lines are gauge poles and black lines are poles of $\lambda$. Regions I-III correspond to the same regions as Figure \ref{fig:mass space}. (Right) Thermal targets for representative dark matter candidates. The black, gray, and red curves represent the parameter space for which the abundance of $\chi$ is in agreement with the observed dark matter energy density, where $\lambda_0$ is set to the lowest possible value such that $m_h \geq m_\chi$. The shaded region above the purple curve is excluded by the BaBar $\gamma^+$ missing energy search \cite{Izaguirre:2013, Essig:2013}. The dashed purple curve is the projected sensitivity of a $\gamma^+$ missing energy search at Belle II using $20 \text{ fb}^{-1}$  \cite{CosmicVisions17, Belle-II:2022cgf} 
    . The shaded region above the solid blue curve is excluded by the NA64 missing energy search for DM with $\alpha_D = 0.1$ \cite{NA64}. The dashed blue and green lines are the projected capabilities for an NA64 electron and muon beam respectively. The red dashed curve is the LDMX projection for a 10\% radiation length tungsten target and an 8 GeV beam presented in \cite{Hearty:2017} , which was scaled up to $10^{16}$ EOT relative to a background study with 4 $\times 10^{14}$ EOT \cite{ldmx}.}
    \label{fig:p1 scalar}
\end{figure}

In principle, we should consider higher P (lower $Q_\phi$) theories, for all of the DM models discussed here. We will make some heuristic arguments here about how high we can feasibly take P for each model. We start by recognizing that, when making order of magnitude estimates, $m_\chi \sim v$, where $v = \sqrt{2}\langle \phi \rangle$, because $3 m_\chi = m_{A'} = g_\phi v$. Then we can note that Majorana masses in a fermion theory are related to $v$ by
\begin{equation}
    m_M  \propto \frac{v^P}{\Lambda_\star^{P-1}} \sim \frac{m_\chi^P}{\Lambda_\star^{P-1}},
\end{equation}
where $\Lambda_\star$ is a cutoff scale where new particles are required to enter the theory. Flipping this around we see that this cutoff scale is given by
\begin{equation}
\Lambda_\star \propto \left(\frac{m_\chi^P}{m_M}\right)^{\frac{1}{P-1}}.
\end{equation}
Now let's begin with MDM where the Majorana mass is of the same order as the DM mass: $m_M \sim m_\chi$. In this case, for any $P>1$ we find that
\begin{equation}
\Lambda_\star \propto m_\chi,
\end{equation}
which is exactly why we cannot have $P > 1$ for MDM. On the other had, when considering PDDM, we can take $m_M$ as low as $\sim 10^{-4} m_\chi$. While the mass splitting is necessarily small for PDDM, it avoids CMB bounds provided the splitting is large enough to deplete the heavier state in the early universe \cite{CarrilloGonzalez:2021}.  In this case we have a cutoff given by
\begin{equation}
\Lambda_\star \propto 10^{\frac{4}{P-1}}m_\chi,
\end{equation}
thus allowing $P>1$ for modest values of $P$.  There is a trade off, however, because as we raise $P$ we also raise the number of new charged particles that need to enter in the UV, which will in turn lower the gauge Landau pole. Modeling the full UV behavior of $P>2$ models is left to future work. In Scalar DM, the gauge symmetry breaking mass term is related to $v$ by
\begin{equation}
    \mu_\chi^2  \propto \frac{v^P}{\Lambda_\star^{P-2}} \sim \frac{m_\chi^P}{\Lambda_\star^{P-2}},
\end{equation}
for which we can take the mass ratio of $\mu_\chi = 10^{-6}m_\chi$. For scalar DM there is no CMB requirement for an appreciable mass splitting since SEDM has $p$-wave annihilations. However, mass splitting around $\sim 10^{-6} m_\chi$ is roughly the minimum size for which direct detection constraints change between the elastic and inelastic case. This leads to the cutoff energy
\begin{equation}
    \Lambda_\star \propto 10^\frac{12}{P-2}m_\chi.
\end{equation}
Thus, we might be able to take $P$ significantly higher for Scalar DM. Again, we leave this for future work. For completeness, however, we will map out the parameter space for a $P=1$ theory of Scalar DM, since this gives rise to a relevant operator. The couplings run the same as in $P \geq 2$ theories with appropriately altered charges, and in Figure \ref{fig:p1 scalar} the $\lambda$ and gauge poles are plotted as contours on the $\alpha_D - \lambda_0$ plane, as well as lines of experimental reach analogous to Figure \ref{fig:exp reach}.

\section{Understanding the Phenomenology of Different Dark Higgs Masses}
\label{app: pheno}
\begin{figure}[h]
    \centering
    \includegraphics[width = 15cm, height = 5cm]{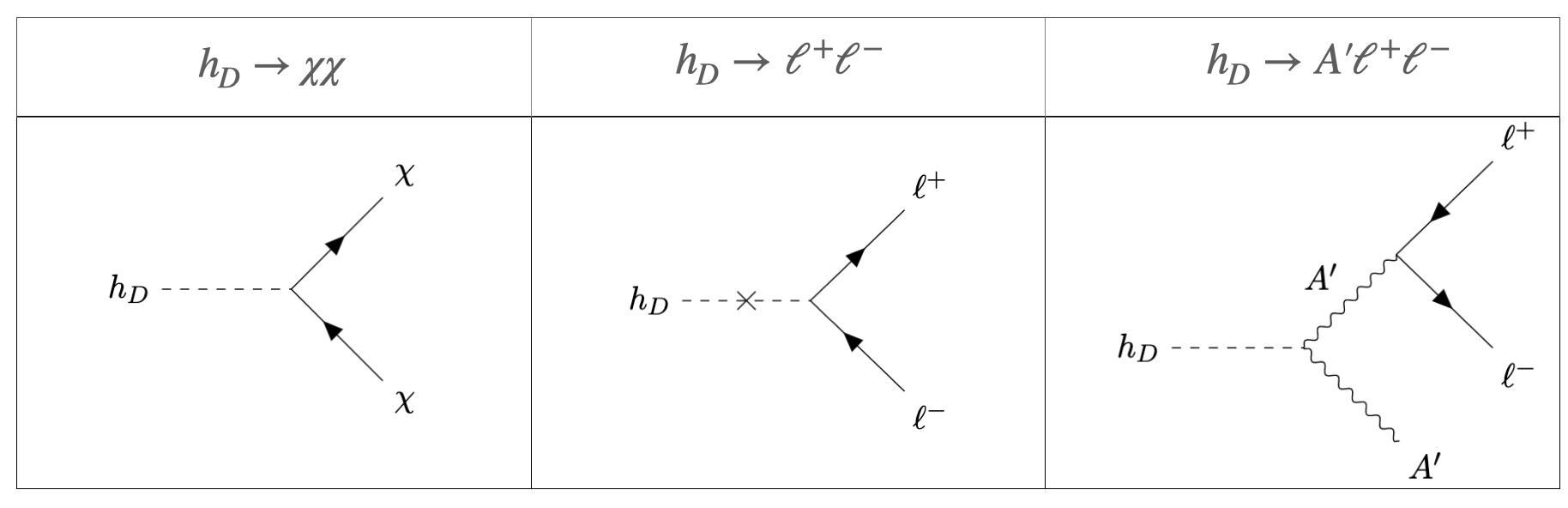}
    \caption{Feynman diagrams for the most prominent dark Higgs decay products.}
    \label{fig:higgs decay}
\end{figure}

Depending on what the dark Higgs to DM mass ratio is, producing a dark Higgs $h_D$ at a collider might have bright decay signatures. When the dark Higgs is less than twice the DM mass, any on shell Higgs produced, for example in a dark higgstrahlung process, will decay to visible particles. Once the dark Higgs is more than twice the DM mass, it will decay almost entirely back to the dark sector. However, it is worth double checking what the rate of $h_D$ decay to DM is compared to $h_D$ decay to SM particles. Below the mass of the dark photon the main competitor will be to SM leptons via Higgs mixing.  Above the mass of the dark photon, $h_D$ can also decay to one dark photon and two SM leptons. The Feynman diagrams for these three decays are shown in  Figure \ref{fig:higgs decay}.

Computing the decay rate of $h_D$ to DM particles is fairly straightforward,
and we will work only with the PDDM $P=1$ case for clarity, though the qualitative results should hold for the others. From there, we can parameterize the Yukawa coupling as $\frac{\delta_{m_\chi}}{\langle \phi \rangle}$, and thus get the following rate
\begin{equation}
\Gamma (h\rightarrow \chi\chi) = \frac{1}{16\pi}\left(\frac{\delta_{m_\chi}}{\langle \phi \rangle}\right)^2m_{h}\left(1 - 4 \frac{m_{\chi}^2}{m_{h}^2}\right)^{\frac{3}{2}} .
\end{equation}
We now compare this to direct decay to SM matter, with muons as a representative example in the Sub-GeV mass range. We will parameterize the dark Higgs to SM Higgs coupling via a mixing angle $\sin{\theta}$. Then, writing the muon to Higgs Yukawa coupling as $\frac{m_\mu}{\langle H \rangle}$, we get the following decay rate
\begin{equation}
\begin{split}
\Gamma (h\rightarrow  \mu^+\mu^-) = \frac{1}{8\pi}(\frac{m_\mu}{\langle H \rangle }\sin{\theta})^2m_h\left(1 - 4 \frac{m_{\mu}^2}{m_h^2}\right)^{\frac{3}{2}} .
\end{split}
\end{equation}
Finally, we compute the decay rate of $h_D$ into $A'$ and an $e^+e^-$ pair via kinetic mixing
\begin{equation}
\begin{split}
\Gamma (h\rightarrow  A'e^+e^-) = \frac{m_h^3g_\phi^2}{m_A^2}\frac{1}{64\pi^3}(e\epsilon)^2 F(\frac{m_A}{m_h}),
\end{split}
\end{equation}
where 
\begin{equation}
    F(\varepsilon) = \varepsilon^2\int_\varepsilon^{(1+\varepsilon^2)/2}dx (x^2 - \varepsilon^2)^{1/2} \frac{(3\varepsilon^4) + 2\varepsilon^2 - 6\varepsilon^2 x + x^2}{(1-2x)^2}
\end{equation}
is an approximate phase space factor in the limit of massless $e^+$ and $e^-$ \cite{Rizzo1980}. With these equations at hand, we can compare the rate of a specific example to see how they compare. For a PDDM theory with $\delta_{m_\chi} = 10^{-4}m_\chi, m_\chi = 0.2 \text{ GeV}, m_h = 1.5 m_A,$ and $\alpha_D = 0.5 (g_\phi \approx 5)$, we get the following rates
\begin{equation}
\begin{split}
\Gamma (h\rightarrow \chi\chi) \approx 1.79 \times 10^{-8} \text{ GeV},
\end{split}
\end{equation}

\begin{equation}
\begin{split}
\Gamma (h\rightarrow  \mu^+\mu^-)  \approx 6.07 \times 10^{-9} \sin^2{\theta} \text{ GeV},
\end{split}
\end{equation}

\begin{equation}
\begin{split}
\Gamma (h\rightarrow  A'e^+e^-) \approx 2.35 \times 10^{-5} \epsilon^2 \text { GeV}.
\end{split}
\end{equation}
Since $\sin ^2{\theta}$ is well constrained to be $\lesssim 10^{-6}$ \cite{Batell:2022}, and $\epsilon^2$ is $\approx 10^{-8}$ for thermal relic PDDM of this mass, the dark Higgs decay really is overwhelmingly invisible in Region III as defined in Section \ref{poles}.

\end{document}